\documentclass{aastex}
\usepackage{emulateapj5, psfig, epsfig}

\newcommand{\msun}{\ensuremath{\,M_\odot}}

\slugcomment{Accepted by The Astronomical Journal}

\shorttitle{Binaries in NGC 288}
\shortauthors{Bellazzini et al.}

\begin{document}


\title{Deep HST-WFPC2 photometry of NGC 288. I.\\
    Binary Systems and Blue Stragglers
    \thanks{Based on observations made with the NASA/ESA Hubble Space
    Telescope at the Space Telescope Science Institute. STScI is operated by the
    Association of Universities for Research in Astronomy, Inc. under NASA
    contract NAS 5-26555. These observations are associated with proposal \#
    GO-6804.}}


\author{Michele Bellazzini, Flavio Fusi Pecci\altaffilmark{2},
Maria Messineo\altaffilmark{3}}
\affil{Osservatorio Astronomico di Bologna, Via Ranzani 1, 40127, Bologna, 
ITALY}
\email{bellazzini@bo.astro.it, flavio@bo.astro.it, 
messineo@strw.leidenuniv.nl}

\author{Lorenzo Monaco} 
\affil{Dip. di Astronomia, Universit\`a di Bologna, Via Ranzani 1, 40127, 
Bologna, ITALY}
\email{s\_monaco@bo.astro.it}

\and

\author{Robert T. Rood}
\affil{Department of Astronomy, University of Virginia, 
        P.O. Box 3818, Charlottesville, VA, 22903-0818 }
\email{rtr@virginia.edu}

\altaffiltext{2}{Stazione Astronomica di Cagliari, Loc. Poggio
dei Pini, Strada 54, 09012 Capoterra (CA), ITALY}
\altaffiltext{3}{presently at Sterrenwach Leiden, Postbus 9513, 2300 RA Leiden, 
The Netherlands}


\begin{abstract}
We present the first results of a deep WFPC2 photometric survey of the
loose galactic globular cluster NGC 288. The fraction of binary
systems is estimated from the color distribution of objects near the
Main Sequence (MS) with a method analogous to that introduced by
\cite{ruba1}. We have unequivocally detected a significant population
of binary systems which has a radial distribution that has been
significantly influenced by mass segregation. In the inner region of
the cluster ($r<1 r_h \simeq 1.6 r_c$) the binary fraction ($f_b$)
lies in the range 0.08--0.38 regardless of the assumed distribution of
mass ratios, $F(q)$. The most probable $f_b$ lies between 0.10 and
0.20 depending on the adopted $F(q)$. On the other hand, in the outer
region ($r\ge 1 r_h $), $f_b$ must be less than 0.10, and the most
likely value is 0.0, independently of the adopted $F(q)$. The detected
population of binaries is dominated by primordial systems.

The specific frequency of Blue Straggler Stars (BSS) is exceptionally
high, suggesting that the BSS production mechanism via binary
evolution can be very efficient. A large population of BSS is possible
even in low density environments if a sufficient reservoir of
primordial binaries is available. The observed distribution of BSS in
the Color Magnitude Diagram is not compatible with a rate of BSS
production which has been constant in time, if it is assumed that
all the BSS are formed by the merging of two stars.

\end{abstract}



\section{Introduction}

Binary systems are the most frequent form in which stars present
themselves, at least in our local neighborhood \citep{duqma}, and it
is generally believed that the same process of stellar birth commonly
takes place in clusters \cite[e.g.,][and references
therein]{math}.  Furthermore, binaries have a significant impact on the
chemical evolution of galaxies and even toy chemical evolution models
can easily show that our Universe would not have been as it is without
them \citep{port00}.

In collisional stellar systems, binaries can play a key role also in
the dynamical evolution. In particular, binaries provide the
gravitational fuel that stops and eventually reverses the process of
core collapse in globular clusters \cite[][and references
therein]{hual,mh97}. Moreover, the evolution of binaries in clusters
can produce peculiar stellar species such as Blue Stragglers (BSS),
Cataclysmic Variables (CV), Low Mass X ray sources (LMXB), millisecond
pulsars (MSP) and possibly sdB
\citep{bai95,port97a,port97b,sgb,maxt01,gre01,gls}.  
Thus, the study of binary
populations in globular clusters can provide powerful constraints both
on dynamical models or on models of formation of exotic objects.

Despite their potential interest, the actual detection of binaries and
estimate of the binary fraction ($f_B$)\footnote{Defined as the ratio
between the number of binary systems and the total number of cluster
members (i.e. binary systems + single stars), see \cite{hual}.} in
globular clusters has eluded the effort of researchers until very
recent times \cite[see][]{hual,gg,itr}, because of the
challenging observational requirements.

There are three main techniques which have been used to detect binary
populations in globular clusters (reviewed by \citep{hual}): (i) radial
velocity variability surveys \citep{lath}, (ii) searchs for eclipsing
variables \citep{mat96a}, and (iii) searches for a Secondary Main
Sequence \cite[SMS, see][]{rw91,tout99} parallel to the normal Main
Sequence (MS) in the Color Magnitude Diagram (CMD).

Methods (i) and (ii) are based on the actual detection of individual
binary systems. They require large amounts of observing time, since
time series are necessary. Method (i) also requires very high
precision radial velocity measures that limit the luminosity range of
the targets and place great demands on the quality and stability of the
instrumental setup. These lead to observational biases and intrinsic
limitations \cite[see \S2.1 and \S2.2 in][]{hual} and ultimately to a
low discovery efficiency. Thus, while these methods provide the only route to 
the physical characterization of individual systems (masses, orbits etc.),
they are maladapted to determining population properties, such as the binary
fraction.

In contrast, method (iii) is statistical in nature and does not need
repeated observations. It is based on the simple fact that any binary
system at the distance of globular clusters is seen as a single star
with a flux equal to the sum of the fluxes of the two components. In
the CMD, systems composed of two equal mass MS stars lie on a sequence
$0.752$ mag brighter than the single stars sequence. If the masses are
not equal the shift away from the MS is smaller and depends
non-linearly on the mass ratio of the two components,
$q={M_2\over{M_1}}\le 1.0$ \cite[see][]{tout99}, where $M_1$ and $M_2$
are the mass of the primary and secondary stars, respectively. 

While the SMS can detect binaries of any orbital period and any
orientation of the orbital plane, it suffers from three
problems:

\begin{itemize}

\item The observational signatures of a genuine binary system and two
blended, but unrelated, single stars are indistinguishable. Thus, any
SMS in a globular cluster is contaminated  to some
degree by blended objects.

\item Exceptionally accurate photometry extending to at least a couple
of magnitudes below the Turn Off point (TO) is required. Any secondary
sequence can be easily obscured by the observational scatter in the
single star sequence. Thus, while hints of a SMS have been found in
many cases \cite[see Tab. 3 in][]{hual}, the only definitive
detections from ground based photometry have been in two very loose
and relatively nearby clusters, E3 \citep{ver96,mcc85} and NGC 288
\citep{bol92}.

\item The distribution of mass ratios must be retained as a free
parameter while estimating $f_b$, although some observational
constraints on this distribution may emerge.

\item SMS is very sensitive to the distribution of the mass ratios. To obtain
an useful constraint on the binary population, the high $q$ part of the
distribution must be significantly populated.

\end{itemize}          

Sufficient photometric accuracy at such faint magnitudes can often be
be achieved with HST observations, and sometimes with large ground
based telescopes in optimal seeing conditions. The correction for
blendings has a very complex behavior and can be made only via
extensive artificial star experiments which precisely mimic the
observations and data reduction.

While the above framework was clearly recognized, e.g., by
\cite{bol92}, the full development and application of a method that
properly accounted for all these effects emerged only with the seminal
work by \cite{ruba1}.  These authors analyzed a large set of exposures
taken with the Planetary Camera of the HST-WFPC2, imaging a field
including the central region of the globular cluster NGC~6752. 
They were able to estimate $f_b$ from the
distribution of deviation in color with respect to the MS ridge line
using a large set of artificial stars experiments to correct for
blendings and considering different possible distributions of the mass
ratios. They found 15\% $\le f_b \le 38$ \% for the sample within the
core radius of NGC~6752 and marginal evidence for $ f_b \le 16$ \%
outside that limit.

We have used a method very similar to that of \cite{ruba1} [RB97] on
HST observations of two partially overlapping WFPC2 fields sampling
the central part of the low density cluster NGC288.  In \S2 we will
describe the observations, data reduction, and artificial star
experiment. In \S3 the method will be described in detail, outlining
the differences with respect to RB97. Results and discussion are
reported in \S4. Once the binary fraction and its spatial properties
are established, we turn to the analysis of the expected products of
the evolution of binary systems, in particular BSS (\S5). Finally we
summarize the results and provide a global description of the status
and the evolution of the binary population in NGC288 (\S6).
Some preliminary results from the observations presented
in this paper have been reported in a recent meeting \citep{bm00}. 

\section{Observations and Data Reduction}

The observations were taken on November 17, 1997 as part of the
GO-6804 program (P.I., F. Fusi Pecci). Two WFPC2 fields were observed:
one with the PC centered on the cluster center (Internal field;
hereafter the {`Int'} field) and one with the WF4 camera partially
overlapping the WF3Int (External field; hereafter the {`Ext'}
field). The approximate boundaries of the observed fields are shown in
Fig.~1. The position of the center of NGC288 \cite[according
to][]{web85} and some characteristic scalelengths are also shown (see
caption). It is interesting to note that the Int field is almost
completely within the half light radius ($r_h$).

\begin{figure*}
\figurenum{1}
\centerline{\psfig{figure=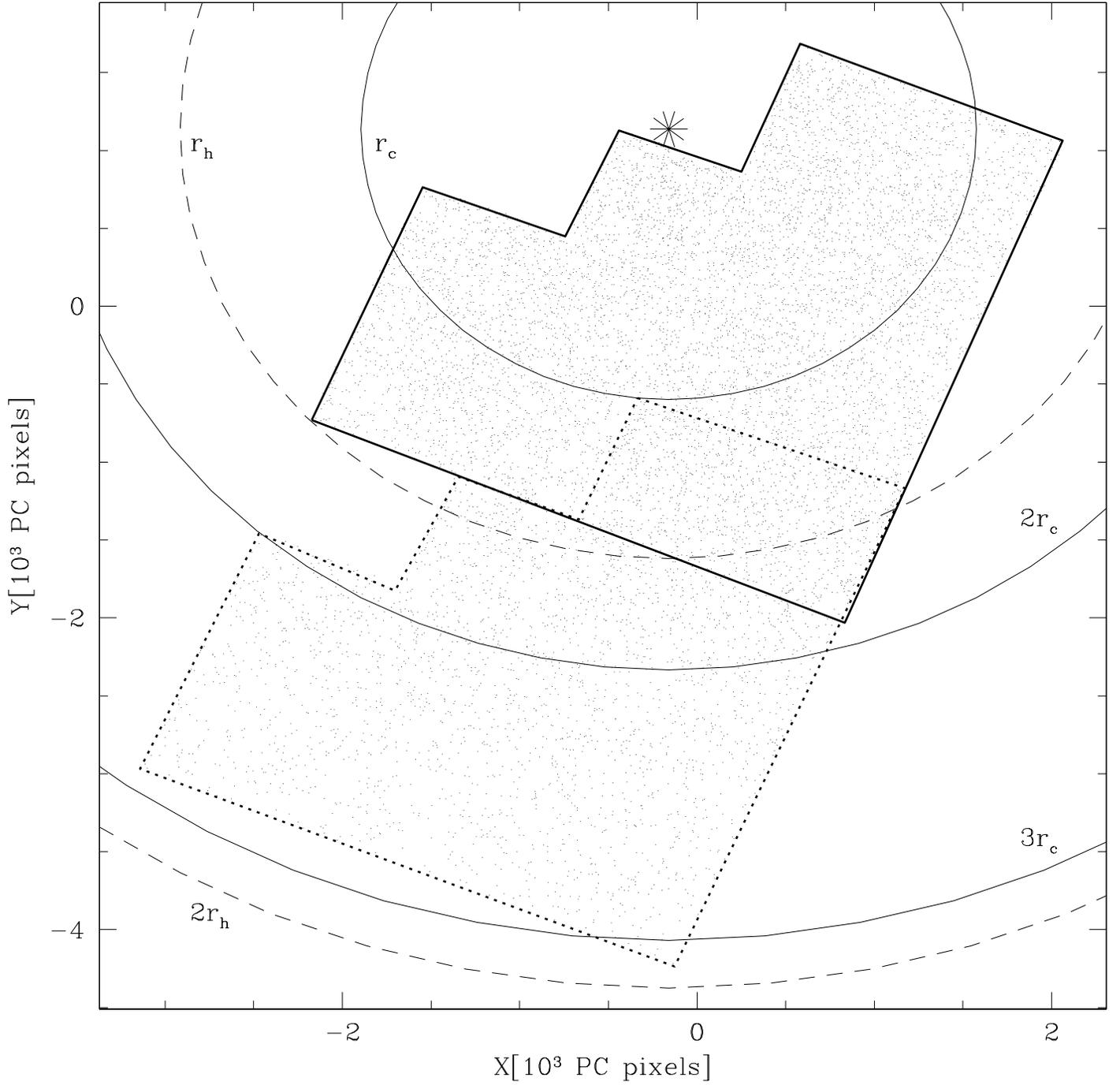}}
\caption{Position in the sky of the observed fields. North is up and East is
right. The heavy continuous line 
is the contour of the Int field and the heavy dotted line is the contour of the
Ext field. The small dots represent the stars included in our final catalog.
The big asterisk is the center of NGC288, according to \citet{web85}. The
continuous circles have radii = 1, 2 and 3 core radii ($r_c$) and the long
dashed circles have radius = 1 and 2 half light radii($r_h$). $r_c$ and $r_h$
are from \citet{trag93}.}
\end{figure*} 

The observational material is described in Table 1. The first column
gives the name of the field, the second column the WFPC2 filter
used, the third column the exposure times, and the fourth column the
number of repeated exposures acquired at the indicated exposure time
and passband.  The repeated exposures have been taken after shifts of
a semi-integer number of pixels in order to minimize the undesired
effects of bad pixels and inaccuracies in the flat field
correction. Furthermore, such shifts guarantee that the same stellar
images are sampled by different subrasters of pixels in each frame,
thus limiting the effects associated with the relative position of the
PSF peak and the pixel/s collecting the core of the stellar image in
the final averaged measure.

\subsection{Data Reduction}   

The data reduction has been performed on the pre-calibrated frames provided by
the STScI. The CTE corrections has been applied according to \citet{whit99}.

In the present work we use mainly the F555W and F814W frames, thus we
will concentrate on these. However the general reduction strategy is
the same independently of the passband. The relative photometry has
been carried out using the PSF-fitting code DoPHOT \citep{dophot},
running on a Compaq Alpha station at the Bologna Observatory.  We
adopted a version of the code with spatially variable PSF and modified
by P. Montegriffo to read real images. A quadratic polynomial has
been adopted to model the spatial variations of the PSF. The
parameters that control the PSF shape have been set in the same way as
\citet{ols98}, who made an accurate analysis of the application of
DoPHOT to WFPC2 images. Since the code provides a classification of
the sources, after each application we retained only the sources
classified as {\em bona fide} stars (types 1, 3 and 7).

The procedure adopted to handle the repeated exposures was:

\begin{enumerate}

\item The single images were shifted to one reference image and combined into a
median frame, cleaned of cosmic rays and other defects. This step has been
performed adopting standard IRAF-STSDAS procedures.

\item The cleaned median frame was searched at a 3$\sigma$ threshold 
with DoPHOT. In this way we obtained a list of sources uncontaminated by
spurious detections associated with cosmic rays or bad pixels.

\item We then reduced all of the original single frames using the so
called ``warmstart'' option of DoPHOT, i.e., forcing the code to fit
{\em only the stars detected in the median frame}. In this way we
avoid the spurious detections (taking advantage of the cleaning of the
median frame), and we obtain a more robust and accurate photometry by
measuring the relative magnitudes of the stars on the single
frames\footnote{It is well known that averaging multiple measures
obtained on single frames provides much more accurate photometry than that
from the reduction of a stacked image when dealing with WFPC2 images. A
clear demonstration has been published by \cite{ruba1}}.

\item The final catalogs were cross-correlated and the multiple measures were
averaged adopting a 2-$\sigma$ clipping algorithm to reject measures in obvious
disagreement with the others measures of the same stars. Only stars with at
least two valid measures in each filter were retained in the final catalog.

\end{enumerate}

The catalogs obtained from frames of different exposure time were
merged by converting the short $t_{\rm exp}$ photometry into the
photometric system of the longer $t_{\rm exp}$ catalog and adopting
merging thresholds that retained the measures with the
highest signal to noise ratios.

The crowding conditions in our images are never critical, thus it was
not difficult to derive robust estimates of the aperture correction at
$0.5$ arcsec apertures with IRAF/PHOT, for both the Int and Ext
fields. Absolute calibrations in the STMAG and Johnsons-Cousins system
have been obtained by applying the relations by
\citet{holtz95}. Small shifts ($\le 0.02$ mag) were applied to the Ext
photometry to transform to the Int photometric system and thus 
obtain a homogeneous sample over the whole observed field.

Since the \citet{holtz95} absolute calibrations are intrinsically
uncertain we compared our photometry to the only other published $V,~I$
photometry of NGC 288 (Rosenberg, et al. 2000). We then
adopted tiny first degree transformations to transform our photometry to
the well established system of \citet{ros00}. We also checked our
final $V$ photometry with that of \citep{berg93} and found
excellent agreement \cite[see][for further details]{bm01}.

The final CMDs for the Int and Ext fields are presented in
Fig.~2. Only the stars with ${\rm error}_V \le 0.1$ mag and ${\rm
error}_I \le 0.1$ mag are presented.  For the Ext sample the
saturation level occurs at $V\sim 16.0$.  Stars brighter than this
limit are present in the field but do not appear in our catalogs.

\begin{figure*}
\figurenum{2}
\centerline{\psfig{figure=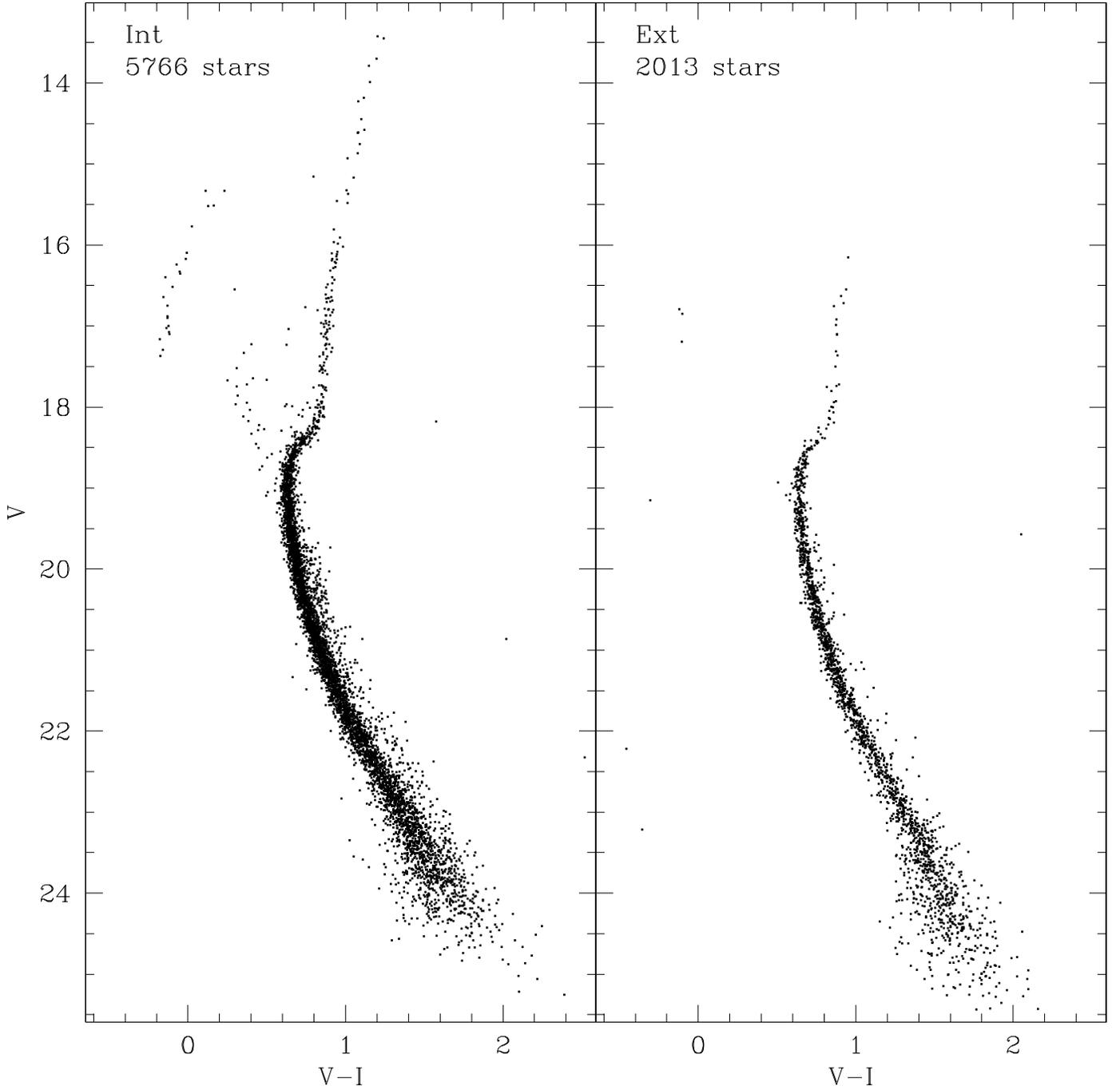}}
\caption{$V,~I$ CMDs of the Int (left panel) and Ext (right panel) fields. Only
the stars with observational errors lower than $0.1$ mag in each filter are
presented in the plots. Stars brighter than $V\sim 16.0$ in the Ext
field have saturated images and are not shown}
\end{figure*} 

We will not comment in detail on the CMDs here. In a companion paper
(Paper II) we will provide a deeper analysis of the evolutionary
sequences and of the Luminosity Function (LF). The relevant points
for the present study are: (a) the quality of the CMDs is excellent,
the average photometric error is $\le 0.02 - 0.03$ mag over the whole
range of magnitudes covered by the diagrams, (b) a well populated SMS
is clearly evident, and (c) a significant population of BSS candidates
is present in the Int sample.

\subsection{Artificial Star Experiments}

Syntheticly reproducing of the complete process of photometric
measurements is the only way to properly characterize of all the
undesired effects associated with observations in a crowded stellar
field. In the present context a large number of artificial star
experiments are crucial. As noted above the effects of photometric
errors plus blending gives a signature quite similar to the widening
of the MS associated with a population of binary systems. To
disentangle the two requires a high degree of accuracy and statistical
significance in any subregion of the observed field.

For each individual subfield (i.e., PC-Int, WF2-Int, WF3-Int, WF4-Int,
PC-Ext, WF2-Ext, WF3-Ext, WF4-Ext) the adopted procedure for the
artificial star experiments, was the following:

\begin{enumerate}

\item A ridge line covering the whole range of magnitudes in the CMD
was obtained by averaging over 0.4 mag boxes and applying
a 2-$\sigma$ clipping algorithm.

\item The $V$ magnitude of artificial stars was randomly extracted
from a $V$ Luminosity Function (LF) modeled to reproduce the observed
LF for bright stars ($V<20$) and to provide large numbers of faint
stars down to below the detection limits of our observations ($20\le V
< 28$). Note that the assumption for the fainter stars is only for
statistical purposes, i.e., to simulate large number of stars in the
range of magnitude where significant losses due to incompleteness are
expected. Furthermore, the actual choice of the LF of the artificial stars is 
not important in the present case since the final estimate of the binary 
fraction is based on stars {\em colors} (see \S3.).
At each extracted $V$ magnitude the correct $I$ magnitude
is determined by interpolation on the cluster ridge line. Thus the
(input) artificial stars lie all on the cluster ridge line on the CMD.

\item It is of the utmost importance that the artificial stars do not
interfere with each other, since in that case the output of the
experiments would be biased by {\em artificial} crowding, not present
in the original frame. To avoid this potentially serious bias we have
divided the frames into grids of cells of known width ($\sim 80$
pixels) and we have randomly positioned {\em only one artificial star
per cell} for each run \cite[a similar procedure has recently been
adopted by RB97, and by][]{manu00}. In addition, we constrain each
artificial star to have minimum distance ($\sim 20$ pixels) from the
edges of the cell. In this way we can control the minimum distance
between adjacent artificial stars. At each run the absolute position
of the grid is randomly changed in a way that, after a large number of
experiments, the stars are uniformly distributed in coordinates
\cite[see also][]{tosi}.

\item The stars were simulated with the DoPHOT model
for the fit, including any spatial variation of the shape of the PSF,
and were added on the original frame including poisson photon
noise. Each star has been added to the $V$ and $I$ median frames and
to all the associated single $V$ and $I$ frames.  The measurement process
has been repeated {\em in the exactly same way as the original
measures and applying the same selection criteria} described in the
previous subsection.

\item The results of each single set of simulations was appended to a file until
the desired total number of simulations was reached. The final result for each
subfield is a list containing the input and output values of positions $(X,~Y)$
and  magnitudes ($V,~I$).

\end{enumerate}

More than $80,000$ artificial stars were produced in each subfield, and the
total number is $> 1,500,000$. The whole procedure was driven by an automated
pipeline taking advantage of the large degree of automation of DoPHOT. A set of
$100,000$ experiments on a subfield was typically completed in $\sim 6$ hours,
running on a Compaq Alpha station.

\begin{figure*}
\figurenum{3}
\centerline{\psfig{figure=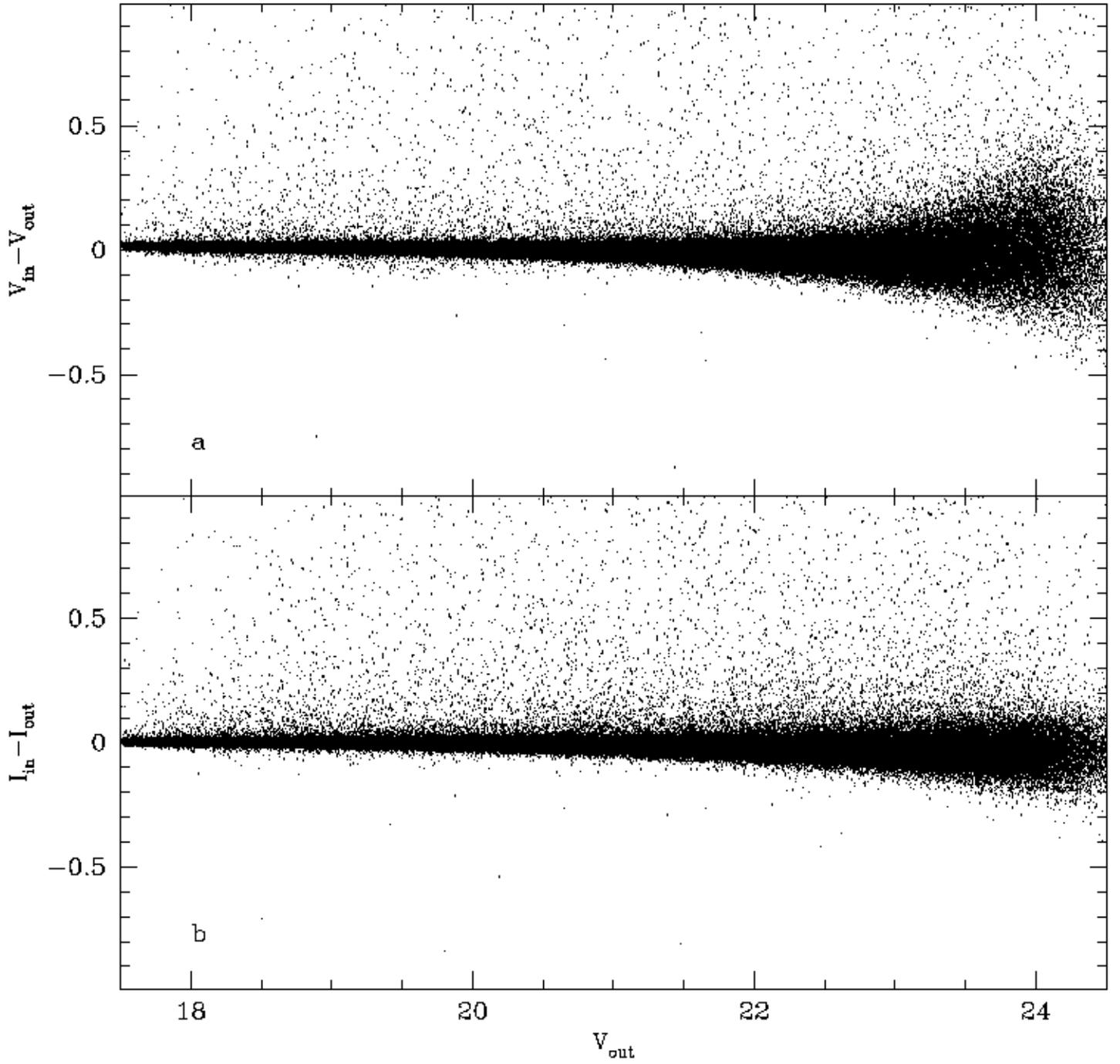}}
\caption{ Differences between the input and output magnitudes of the artificial 
stars versus output $V$ magnitudes for the $V$ [panel ({\em a})] and $I$ 
[panel ({\em b}] passbands, for the WF3int subfields shown as an exemple.}
\end{figure*} 

In Fig.~3 the differences between input magnitudes and output magnitudes are
reported as a function of input magnitude for the artificial stars simulated and
recovered in the WF3Int subfield, chosen as an example case 
(panel ({\em a}): $V_{\rm in}-V_{\rm out}$ vs $V_{\rm out}$, 
panel ({\em b}): $I_{\rm in}-I_{\rm out}$ vs $V_{\rm out}$). 182,441
stars were simulated in this case and 145,608 were recovered.
The plots shows that our measures are very accurate: for instance, the average 
$\sigma(V_{\rm in}-V_{\rm out})$ and $\sigma(I_{\rm in}-I_{\rm out})$
are both $\le 0.05$ mag in for  $V_{\rm out}< 23$.

The distributions of magnitude differences are not symmetrical: there
is a almost uniform cloud of stars in the upper half of each plot that
has no counterpart in the lower half region. These stars have been
recovered with a significantly brighter magnitude than that assigned
in input---they are the artificial stars that blended with a real stars
of similar (or larger) brightness.

\begin{figure*}
\figurenum{4}
\centerline{\psfig{figure=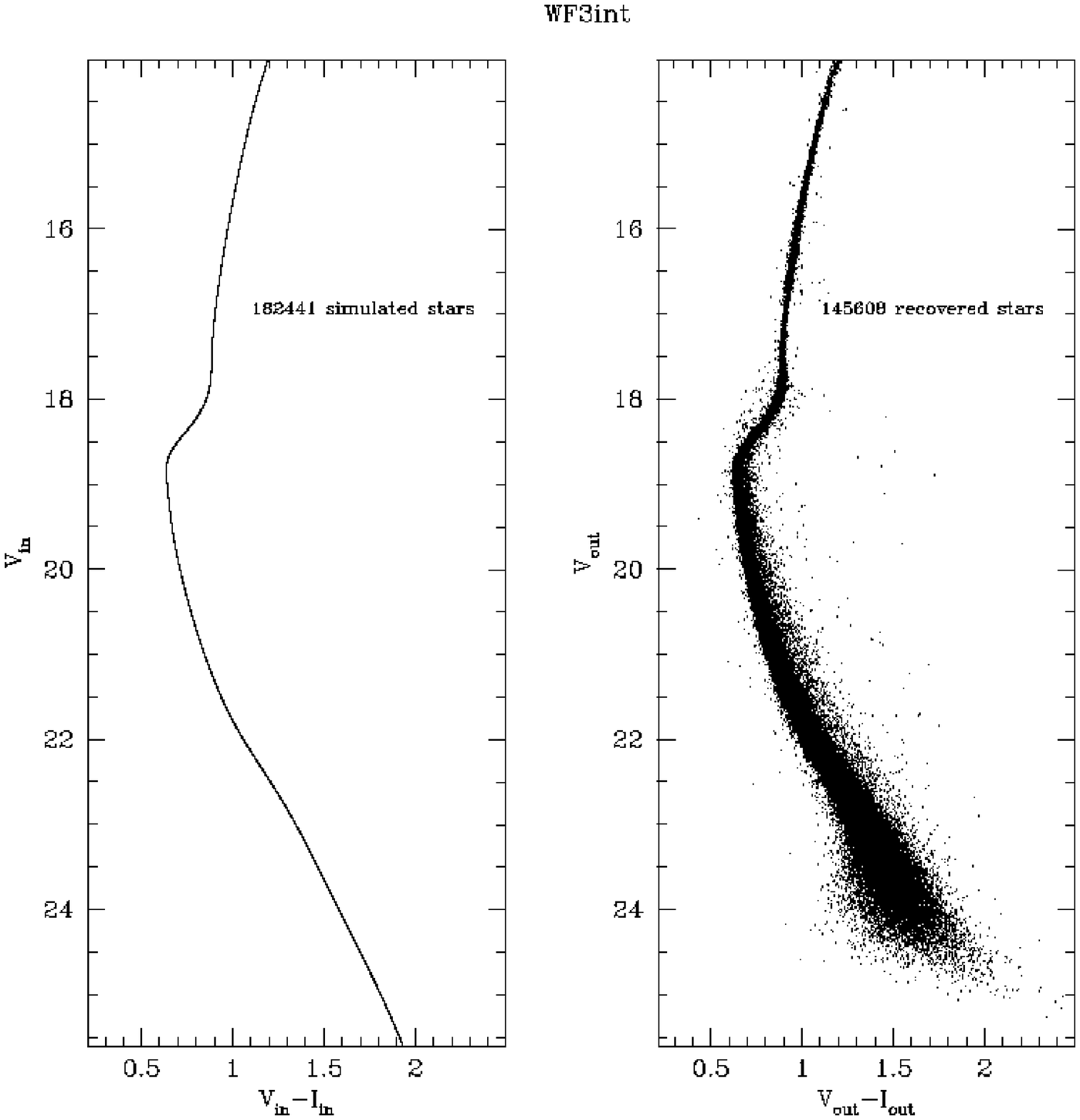}}
\caption{ The CMD for the artificial stars obtained with input magnitudes (left
panel) and output magnitudes (right panel). As an example, the results for the 
WF3int subfields are shown. The left diagram is mapped into the right diagram by
the processes of observation and measure. Note the widening of the base of the
RGB at $V>17$---this is the junction between the long and intermediate
photometry. The stars just fainter than this limit are the highest S/N
sources extracted from the long exposure frames, while the stars just
brighter than this limit are lowest S/N sources of those extracted
from the intermediate exposure frames. It is also interesting to note that the
SMS is well populated by the numerous blendings occurred.}
\end{figure*} 

Fig.~4 shows the input (left panel) and output (right panel) CMD of the
same set of artificial stars. Note the excellent agreement between the observed
(Fig.~2) and simulated (Fig.~4, right panels) CMDs.

It is important to realize that the output CMDs demonstrate the effect
of the observation and data reduction processes on the input stars,
thus it can be thought as modeling a function that maps the {\em true}
distribution of stars in the $(V,~V-I)$ plane into the {\em
observed}\footnote{Where the process of observation include the whole
process, both photon collection and data reduction.} version of the
same distribution. Stars in the right panel of Fig.~4 have been
subjected to all the effects that moved real stars from their {\em
true} position in the $(V,~V-I)$ plane to their actually observed one.

\begin{figure*}
\figurenum{5}
\centerline{\psfig{figure=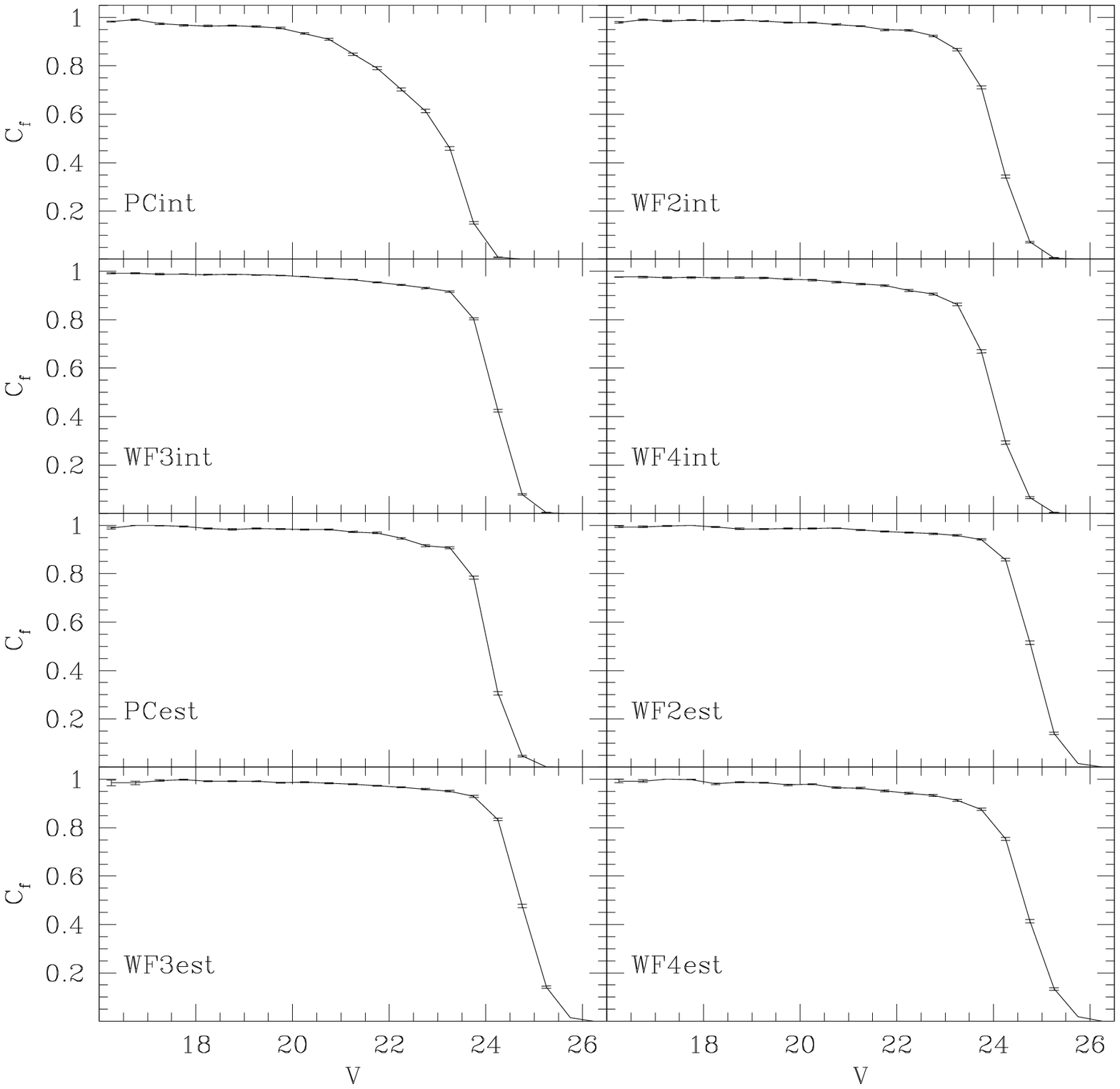}}
\caption{ Completeness factors as a function of $V$ magnitude for the 8 
subfields.}
\end{figure*} 

As a check on the ability to recover stars of the whole procedure we
show in Fig.~5 the completeness factors [$C_f=N_{\rm recovered}/N_{\rm
added}$] as a function of $V$ for each of the eight subfields
corresponding to the different chips of the {\em Int} and {\em Ext}
fields. Note that the continuous line is not a fit to the data but is
simply the line {\em connecting} them. The very low noise of the
observed curves is a spin-off of the very large number of artificial
star experiments performed. With the sole exception of the PC-Int
subfield the completeness is quite similar everywhere and is larger
than $80$ \% at $V\le 24$.  The worse performance of the PC camera
are mainly due to a brighter limiting magnitude at fixed $t_{\rm exp}$
with respect to WFs, because of the smaller pixel scale. We tested the
possible crowding variation within each subfield by comparing the
$C_f(V)$ curves obtained in different quadrants. In all cases the
maximum quadrant to quadrant differences were $< 2$--4\% over the
whole magnitude range. Thus, we concluded that the crowding conditions
are very similar everywhere within each single subfield.

\section{The Method}

Our method for estimating $f_b$ is based on the comparison of the
distributions of the color deviation from the Main Sequence Ridge Line
(MSRL) of the real stars and of appropriate sets of artificial stars
originally placed on the same MSRL. This is exactly as devised by
RB97. We can imagine the effect of observation and data reduction on a
stars as a sum of pulls that move it from the {\em true} place it
would occupy in the CMD: photometric error provides a random pull
while blending or the occurrence of a real binary system pulls the
star systematically toward redder and brighter positions.  The
artificial star experiments optimally characterize the effects of
photometric errors and blendings associated with the considered
observation but do not include any pull due to the occurrence of a
real binary. Thus, their distribution of color deviations from the
MSRL should be different from that of a sample of real stars if binary
systems are present. To estimate $f_b$, we produced additional
artificial sets of MS artificial stars introducing a given fraction of
binary systems, and then we compare the distribution of color deviations
of the synthetic population to that of real stars, searching for the
$f_b$ providing the best match to observations.

We apply the method to the MS in the range $20\le V \le 23$, were the
SMS signal is most easily detected. For $V<20$ the MS is nearly
vertical in the CMD and no significant color deviation can be
detected. For $V>23$ the widening of the sequence due to photometric
errors is too large to distinguish it from that
introduced by a population of binary systems.

\begin{figure*}
\figurenum{6}
\centerline{\psfig{figure=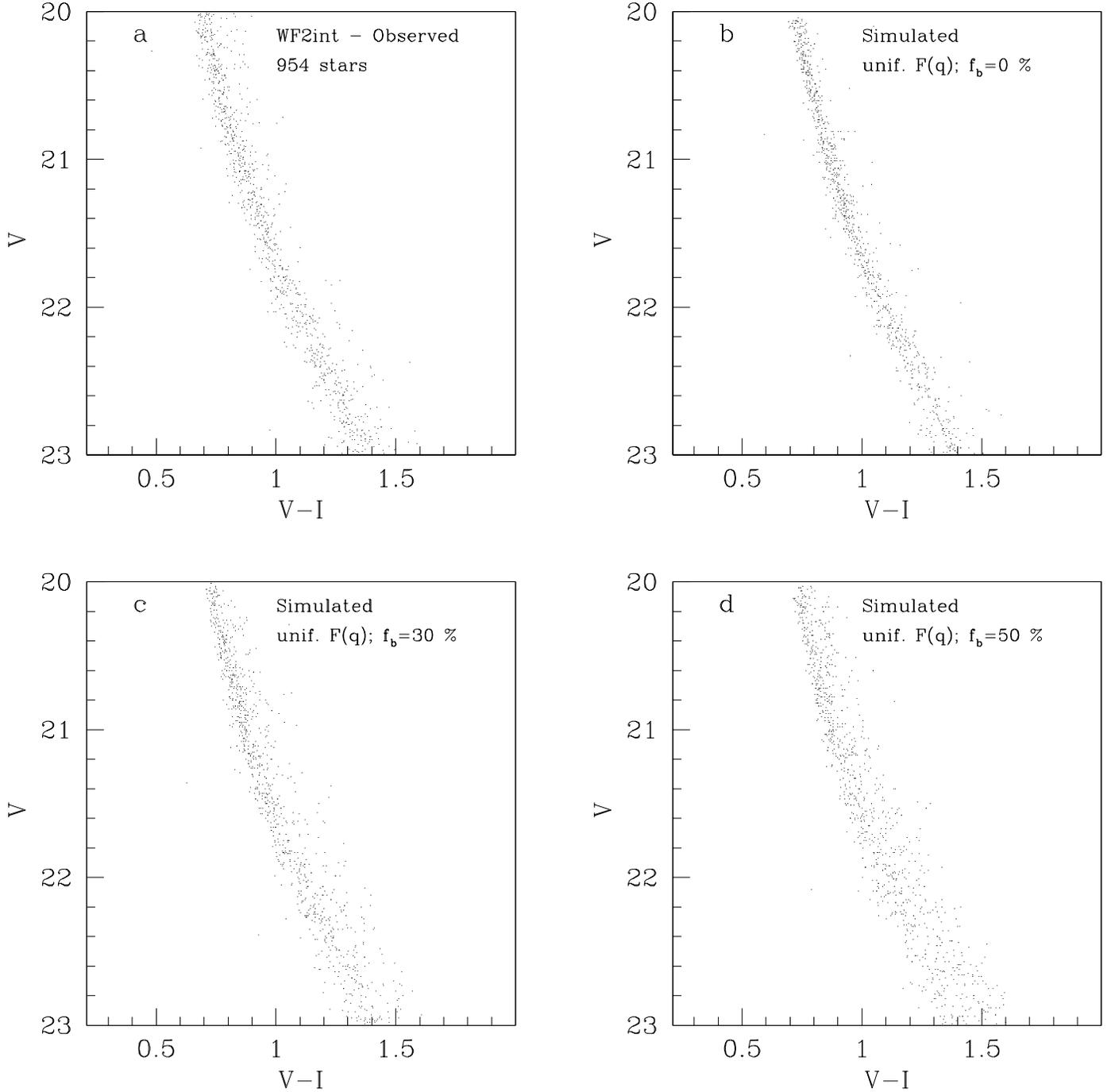}}
\caption{CMDs in the range $20\le V \le 23$
for (a) the sample of real stars observed into the WF2int subfield, 
(b) a synthetic catalog with $f_b=0$, (c) a
synthetic catalog with $f_b=30 \%$ and uniform $F(q)$ and (d) a synthetic
catalog with $f_b=50 \%$ and uniform $F(q)$. All the simulated CMDs contains
the same number of stars as the observed sample and are extracted from the set
of artificial stars simulated on the WF2int chip. Note that each of the
simulations shown is just one of the infinite possible realizations of the CMD
with the given $F(q)$ and $f_b$.}
\end{figure*}

The general scheme adopted is the following:

\begin{enumerate}

\item Stars are randomly generated from an appropriate mass function.  
The associated $V$ magnitudes are obtained from an appropriate 
Mass vs. Luminosity relation obtained from theoretical isochrones. 
The corresponding $I$ magnitudes are obtained from the MSRL.

\item A given fraction of the generated stars is assumed to be the
primary of a binary system; the secondary mass is assigned by randomly drawing
from a mass ratio distribution $F(q)$.
>From the extracted $q$ the mass and the $V$ and $I$ magnitudes are
assigned to the secondary star, the $V$ and $I$ fluxes of the primary
and secondary are summed and the final $V$ and $I$ magnitude of the
unresolved binary system is obtained. The final product
of this process is a list of $V$ and $I$ magnitudes, a fraction of which has
been modified by the addition of the flux by a companion. The total number
of entries of the list is equal to the size of the sample of real
stars we have to compare with.

\item We associate to each entry in the list described above an
artificial star having $V_{\rm in}$ within $\pm 0.03$ mag and that has
been successfully recovered.  Thus for each entry, either representing
a single star or a binary system, we also obtain the corresponding
$V_{\rm out}$ and $I_{\rm out}$ that takes into accounts all the
possible pulls due to observational errors and eventual blendings. 
The final product is a list of synthetic stars
with the {\em same characteristics} of the real stars, in which a
known fraction of binary systems has been introduced.  
Note that each real star has a
counterpart in the synthetic catalog which in turn has a final $V$ and
$I$ from an artificial star that was recovered in the {\em same
region} of the field were the real star is. 
Thus local variations of the crowding are taken into account and
the whole process is done independently for each subfield.  In Fig.~6
we show as an example, the CMD in the range $20\le V \le 23$ for (a)
the WF2int sample of real stars, (b) a synthetic catalog with $f_b=0$,
(c) a synthetic catalog with $f_b=30$ \% and uniform $F(q)$ and (d) a
synthetic catalog with $f_b=50$ \% and uniform $F(q)$ (see caption for
further details).

\item For each fixed $f_b$ and $F(q)$, 100 synthetic catalogs are generated.
Each of these is one of the infinite number of possible random realizations of
a sample having the same characteristics as the observed one and including
a fraction of binary systems $f_b$ whose mass ratios are extracted from $F(q)$.
The distribution of color deviations from the MSRL of {\em each} of these
synthetic samples is compared to that of real stars by means of a function of
merit and the {\em average degree of agreement} as well as the 
{\em dispersion in the degree of agreement} is finally determined.

\item The whole process is repeated for a wide grid of binary fractions and
$F(q)$ distributions, and finally a probability curve as a function of $f_b$ is
produced for each assumed $F(q)$ (cf., Fig.~6 of RB97).

\end{enumerate}

We now examine in more detail some of the single steps.

\subsection{The Mass Function}

Since our original aims included placing constraints on the
distribution of mass ratios, we preferred to produce our synthetic
samples by extracting masses instead of luminosities or magnitudes, as
done by RB97. Our tests showed that useful constraint on $F(q)$ from the
SMS morphology can be derived only if the {\em true} binary fraction
in the sample under consideration is larger than $ 30$ \%.  We adopted
a Mass Generating Function (MGF) similar to that presented by
\citet{jt99}, adjusted to fit the observed mass function of NGC~288.
The MGF is just a numerical algorithm that maps random numbers between
0 and 1 into a given mass distribution, the relevant characteristic
being its ability to reproduce a realistic distribution.  In Fig.~7
the Luminosity Function (LF) derived from the adopted MGF (through a
proper $M_V$ vs. Mass relation, described below) is compared to the
completeness corrected LFs from the observed fields. In this case the
samples from the WF cameras have been merged into the WF-Int and
WF-Ext samples, while the PC-Int and PC-Ext LFs are shown
separately. All the reported LFs have been normalized to approximately
match each other in the range $18.5<V<20.5$. The LF from the
Generating Function clearly provides a good representation of the
observed LFs.

\begin{figure*}
\figurenum{7}
\centerline{\psfig{figure=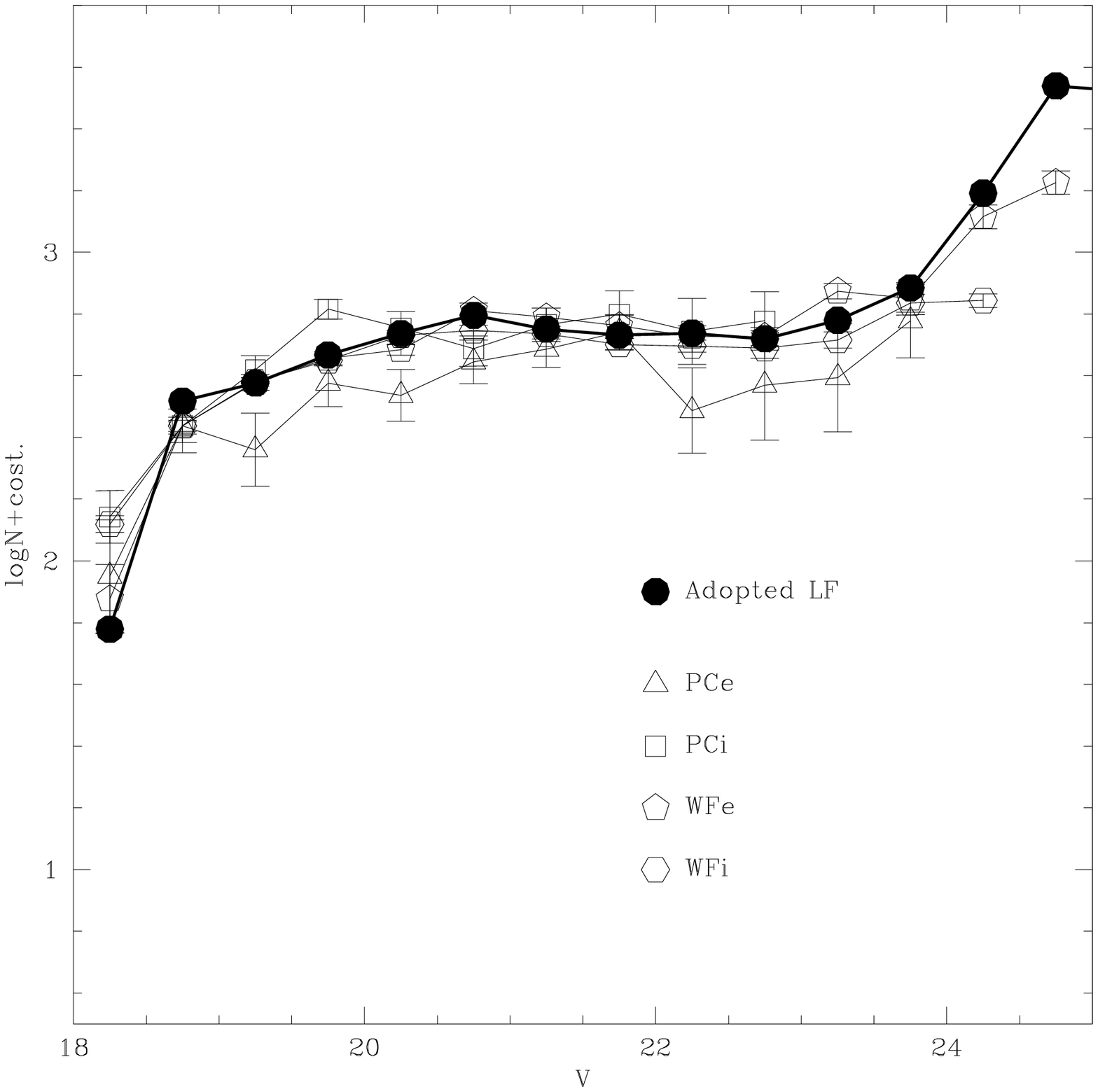}}
\caption{The LF derived from the Mass Generating Function (MGF; large
full dots) compared to the completeness corrected LF for the PC-Int
(open squares), the PC-Ext (open triangles), the combined WF-Ext
cameras (open pentagons) and the WF-Ext cameras (open exagons). All
the LF's are normalized to approximately match in the range
$18.5<V<20.5$. The MGF luminosity function is in good agreement with
the observed ones.}
\end{figure*} 

To convert masses (in solar mass units) into $V$ magnitudes we utilized
theoretical models. In particular, a $Z=0.002$ isochrone from the recent set by
\citet{cas00} provides an excellent fit to the observations once corrected for
distance and reddening [$(M-m)_0=14.73$ and $E(B-V)=0.03$, from \citet{fer99a}], 
as shown in Fig.~8. We use the fitted isochrone to derive the $Mass(V)$ 
function shown in Fig.~9, obtained by cubic spline interpolation.

\begin{figure*}
\figurenum{8}
\centerline{\psfig{figure=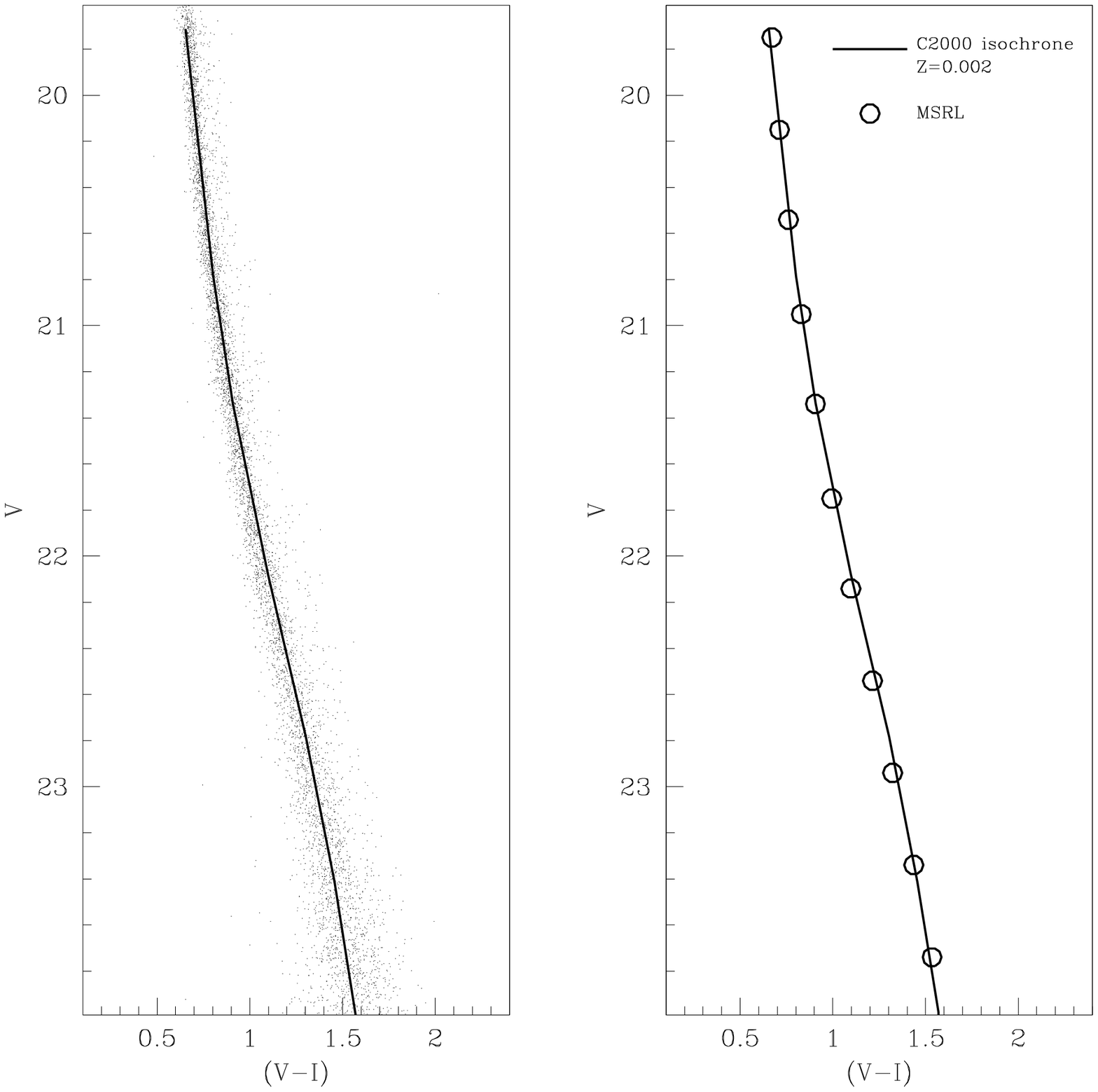}}
\caption{Comparisons between the C2000 isochrone adopted to convert masses to $V$
magnitudes and the observations. In the left panel the isochrone (continuous
line) is superposed to the observed CMD. In the right panel it is superposed
to the MSRL (large open dots).}
\end{figure*} 

It should be noted that the only use of stellar models in the whole
process is the conversion of extracted masses into $V$
magnitudes. Uncertainties intrinsic to the adopted model or associated
with small errors in the fit would have some (moderate) impact only on
the assumed $F(q)$ distributions. The actual $f_b$ estimates are based
on color deviations from the MSRL and are largely independent from the
assumed $Mass(V)$ function.

\begin{figure*}
\figurenum{9}
\centerline{\psfig{figure=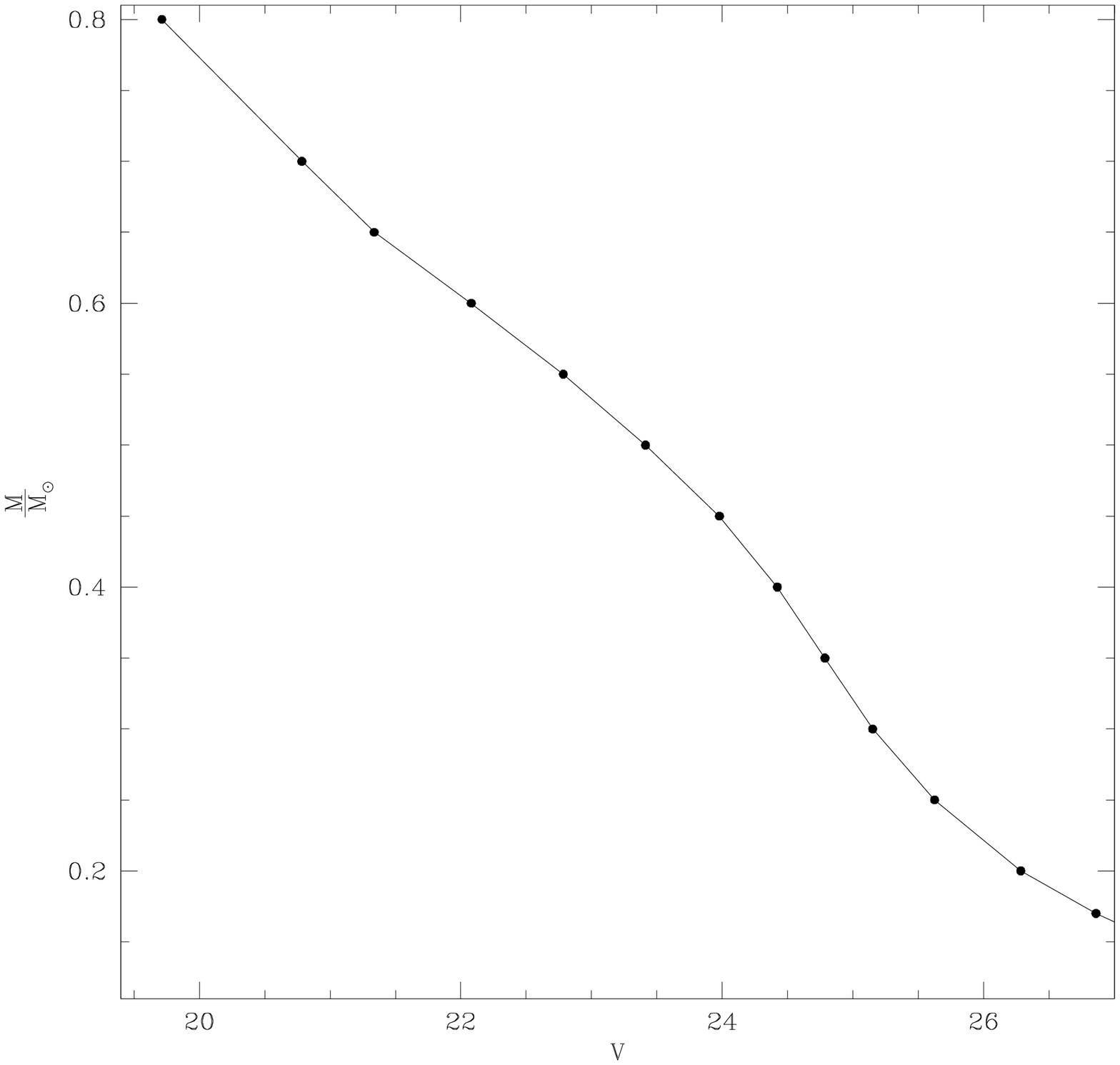}}
\caption{The adopted Mass - $V$ magnitude relation as derived from the
theoretical isochrone presented in Fig.~8.}
\end{figure*}

\subsection{The Mass Ratio Distributions}

Significant samples of local field binaries have been studied by various authors
\citep{trim74,trim86,egg89,duqma}, reaching a reasonable agreement on the
general form of the distribution of mass ratios $F(q)$. 
The local $F(q)$ shows a shallow peak at
$q\sim 0.2$--0.3. The reality of a possible second peak at $q\sim 1$ has been
questioned since it may be produced by selection effects. \citet{tout91} has
shown that, if all of the selection effects are taken into account, the observed
$F(q)$ is naturally reproduced by extracting secondary stars from the observed
Initial Mass Function, i.e. what would be expected by random associations
between stars (hereafter we will refer to this kind of distribution as to the
{\em Natural} $F(q)$). 

There is neither observational constraint on the form of $F(q)$ in
globular clusters, nor theoretical arguments suggesting that the
original $F(q)$ in a globular has to be different from that of the
field. However a difference in the formation of primordial binaries due
to the higher environmental density cannot be excluded.  On the other
hand it is expected that a binary population in a globular may be
subjected to significant evolution through (a) ionization of soft
systems, (b) binary-single star and binary-binary interactions
leading to star exchanges and, eventually, (c) tidal captures and/or
merging of the two components \cite[see][and references
therein]{hual,bai95}.  There is no reason to think that the processes
leading to the disruption or merging of binary systems have some
influence on $F(q)$, and the same conclusion is valid for the tidal
capture phenomenon which is efficient only in extremely dense
environments.  Conversely, any exchange reaction would tend to
preserve the most massive components in the bound system and to eject
the lighter ones, thus moving mass ratios toward unity.

Given this framework, the only possible approach is to keep $F(q)$ as
a sort of free parameter, by measuring $f_b$ under the assumption of
different ``toy models'' for $F(q)$, as done by RB97. While in
principle $0<q\le 1$, this is not necessarily the case while dealing
with the SMS technique, since there is a fundamental limit to the mass
of secondary stars that can contribute additional flux to their primary,
i.e., the stellar mass limit $M\sim 0.08 M_{\odot}$. This limit,
coupled with the maximum mass in the MS ($M\sim 0.8 M_{\odot}$),
sets a lower limit in the range of $q$. All of our model $F(q)$
include this limit to ensure that we remain in the star/star regime
and avoid star/brown dwarf
and star/planet systems\footnote{It is important to recall that the SMS method
is based on the photometric properties of binary systems in which both members
are MS stars, and it is completely insensitive to binary systems including
other stellar species (brown dwarfs, neutron stars etc.).}.

In the choice of the possible $F(q)$ to adopt as test cases we have
been driven by two obvious but opposite requirements, to cover the
widest portion of the parameter space and to avoid an infinite amount
of computing. Thus we considered toy models covering somewhat extreme
cases with $F(q)$ peaking toward the extremes of the $q$ range, an
intermediate one (uniform $F(q)$), and the most realistic model one
can simply conceive, a natural distribution.

Fig.~10 shows the normalized histograms of 100,000 binary systems
randomly drawn from each of the adopted $F(q)$ distributions. The
lower limit $q>0.1037$ has been applied to each of them.  The
distribution Peaked at High Mass Ratios (PHMR; Fig.~10, panel b) has
been produced with a generating function of the form $GF(q)=\sqrt{x}$,
where $x$ is a random number within 0 and 1. For the distribution
Peaked at Low Mass Ratios (PLMR; Fig.~10, panel c) with
$GF(q)=1.-\sqrt{x}$ was adopted, The Natural distribution has been
obtained by extracting the mass of both the primary and
secondary component of the binary systems from the MGF.

RB97 modeled $F(q)$ through distributions of the $V$ magnitude ratio
in the form $V_2=V_1/R^{\xi}$, where R is a random number between 0 and
1 and the exponent $\xi$ determines the form of the distribution. Six
cases where considered, $\xi=0,~0.125,~0.250,~0.5,~1.0,~2.0$. The case
$\xi=0$ is trivial (all binaries have $q=1$). The $F(q)$ associated
with $\xi=0.125$ is very similar to our PHMR, while for $\xi=0.25$
$F(q)$ nearly corresponds to a uniform distribution.  The cases
$\xi=0.5,~1.0,~2.0$ all correspond to $F(q)$ peaked at low mass ratios,
with different importance of the peaks. A lower limit to $q$ similar
to ours was achieved by adopting a faint limit for the magnitude
of the secondary. In the end, the range of mass ratio distributions
sampled in the present analysis is similar to that considered by RB97.

\begin{figure*}
\figurenum{10}
\centerline{\psfig{figure=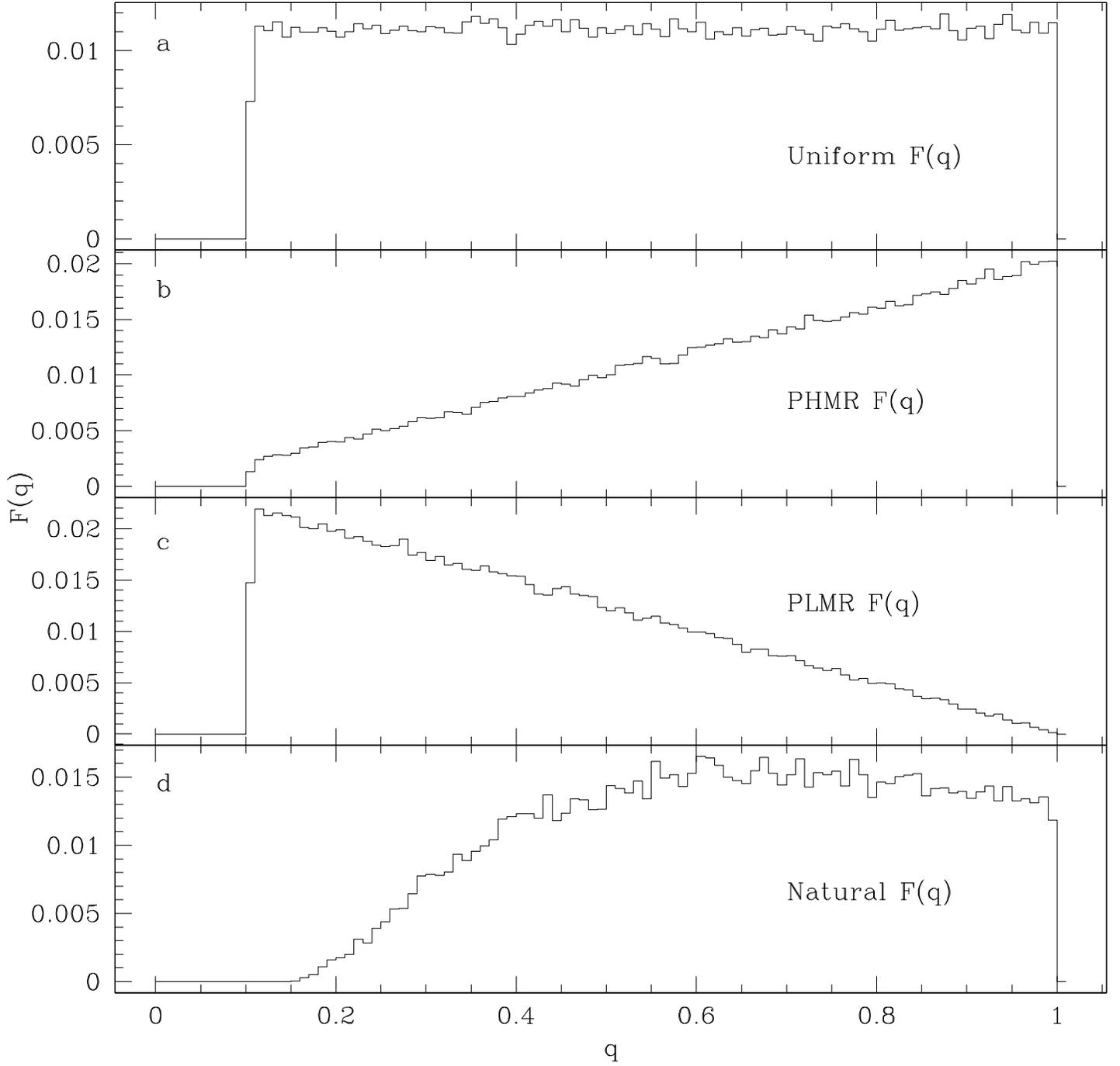}}
\caption{Normalized histograms of 100,000 binary systems randomly drawn from
 the adopted $F(q)$ distributions. Panel a: uniform distribution; panel
b: Peaked at High Mass Ratios (PHMR) distribution; panel c: Peaked at Low 
Mass Ratios (PLMR) distribution; panel d: Natural distribution. The limit
$q>0.1037$ has been applied to all the distributions to exclude stars/brown
dwarf systems from the estimate of the total binary fraction.}
\end{figure*}

\subsection{The Estimate of $f_b$}

We compared the red sides of the $\Delta (V-I)$ distributions of each
synthetic sample with the observed distributions by means of a
Kolmogorov-Smirnov (KS) test as done by RB97. We sample the range of
possible binary fractions at $5$ \% steps
i.e., $f_b=0.05,~0.10,~0.15,~0.20,...,~0.50$. RB97 adopted a $1$ \%
step. For each fixed $f_b$ and $F(q)$ we produced and compared 100
synthetic samples instead of 1000 as done by RB97. The latter two
choices allowed a significant saving in computation time without any
significant reduction of the accuracy of our estimates. In fact our
tests demonstrated that in the present case (a), the $f_b$ with the
highest probability of being compatible with the observation is easily
picked out within the uncertainties with a $5$ \% step, and for case (b) the
mean probability associated with a $[f_b,~F(q)]$ pair as estimated
from 1000 simulations differs by negligible amounts from that
estimated with 100 simulations.

\subsection{Technical Comments}

The present paper reports the final consequences of an extensive set of
tests carried on to optimize the method and to try to extract all the
possible information from the analysis of a well observed SMS. The
results of such tests oriented our choices and drove us to a step by
step definition of the final method. It is interesting to note that in
many cases an alternative to an approach adopted by RB97 for some
specific problem turned out to be unfruitful,
forcing us to follow their path. This may indicate that the basic
characteristics of the method are indeed best suited for the SMS
technique.  A detailed description of the many tests we have
performed is beyond the scope of the present paper. We plan to present
the most interesting results in another paper (Monaco et al., in
preparation). However, there are several points worthy of comment here.

While it is valid to say that we adopted the RB97 method for the
estimating the binary fraction, there are some differences 
in detail. For example, 1) we used a
$Mass(V)$ function and 2) we adopted somewhat different $F(q)$.
In the end, we felt it
worthwhile to describe our method completely.

It is also important to note that the application of the method
requires a very high accuracy in each step of the procedure. For
instance, RB97 found small shifts in $\Delta (V-I)$ introduced by the
artificial star experimental procedure. They were forced to adopt a
MSRL for the artificial stars which was different from the observed
MSRL. Despite of the different pipelines and reduction package used,
the same problem occurred in our study. For our `fix' we felt that it
was a safer choice to adopt a different MSRL for each subset of
artificial stars (i.e., PC-Int, WF2-Int, etc.).  We verified that if
the effect is ignored a significant fraction of the signal associated
with the SMS is lost.

Furthermore, the final MSRLs of each set of artificial stars is based
on $> 50,000$ stars in the relevant range of magnitudes. Thus the
derived MSRL is statistically a very robust representation of the mean
locus of the artificial stars. This is not necessarily the case for
the observed sample. While a robust MSRL can be derived for the {\em
total sample}, tiny camera to camera differences in the
photometry\footnote{This is particularly true for HST-WFPC2
observations, since, for instance, very small misestimates of the sky
level around faint stars may result in significant errors in the final
photometry, see \citet{dol00b} and references therein} can be present,
hence (a) the {\em global} MSRL may not be an optimal representation
of the CMD of each single subfield and (b) the number of stars in each
single subfield may not be sufficient to obtain a ``local'' MSRL
sufficiently good for the present application.
Since case (a) produces a small artificial (and symmetrical) widening
of the observed $\Delta (V-I)$ distributions, fine adjustments are
necessary to obtain consistent comparisons between observed and
synthetic $\Delta (V-I)$.  Otherwise, the relevant distorsion of the
distributions (i.e., those associated with the red tail) would be
smoothed by spurious global discrepancies. We regard this sensitivity
to small inaccuracies as the major drawback of the SMS technique as
developed by RB97 and in the present work.  However, it is important
to stress that if the described adjustments are neglected the fit to the
observed data is worse but the final {\em best fit} $f_b$ estimate is unchanged.

\section{The Binary Fraction of NGC288}

In Fig.~11 [panel ({\em a})] we compare the cumulative distributions of 
color deviations
from the MSRL ($\Delta (V-I)_{MSRL}$) of the whole observed sample to
a randomly extracted set of 70,000 artificial stars without any binary
systems added (i.e., $f_b=0.0$).  It is immediately evident that the
observed distribution is significantly more skewed toward large color
deviations. A KS test shows that the probability that the two
distributions are extracted from the same parent population is
significantly lower than $10^{-6}$. From this simple test, performing
a direct comparison between {\em observed} and {\em artificial} stars,
we obtain a first very important result: {\em the observed
distribution of $\Delta (V-I)_{MSRL}$ in NGC 288 cannot be produced by
a stellar population made only of single stars}, i.e., {the binary
fraction of the cluster is not null, $f_b>0.0$}. Note that the same
conclusion is obtained even if subsamples of artificial stars of the same
size of the observed one are considered.

Since we have demonstrated that $f_b>0.0$ we can now turn to the
actual determination of $f_b$. This goal is obtained through
comparisons similar to that presented in panel ({\em a}) of
Fig.~11. The observed sample is now compared with subsamples of
artificial stars with varying fractions of binary systems added,
according to the prescriptions described in \S3 (see \S3.3, in
particular). One example of a good reproduction of the observed
$\Delta (V-I)_{MSRL}$ distribution is shown in panel ({\em b}) of
Fig.~11. This example was obtained assuming a binary fraction of $15$ \%
and a uniform $F(q)$. According to a KS test, the probability that the two 
compared distributions are drawn from the same parent population is $P=84$ \%.

\begin{figure*}
\figurenum{11}
\centerline{\psfig{figure=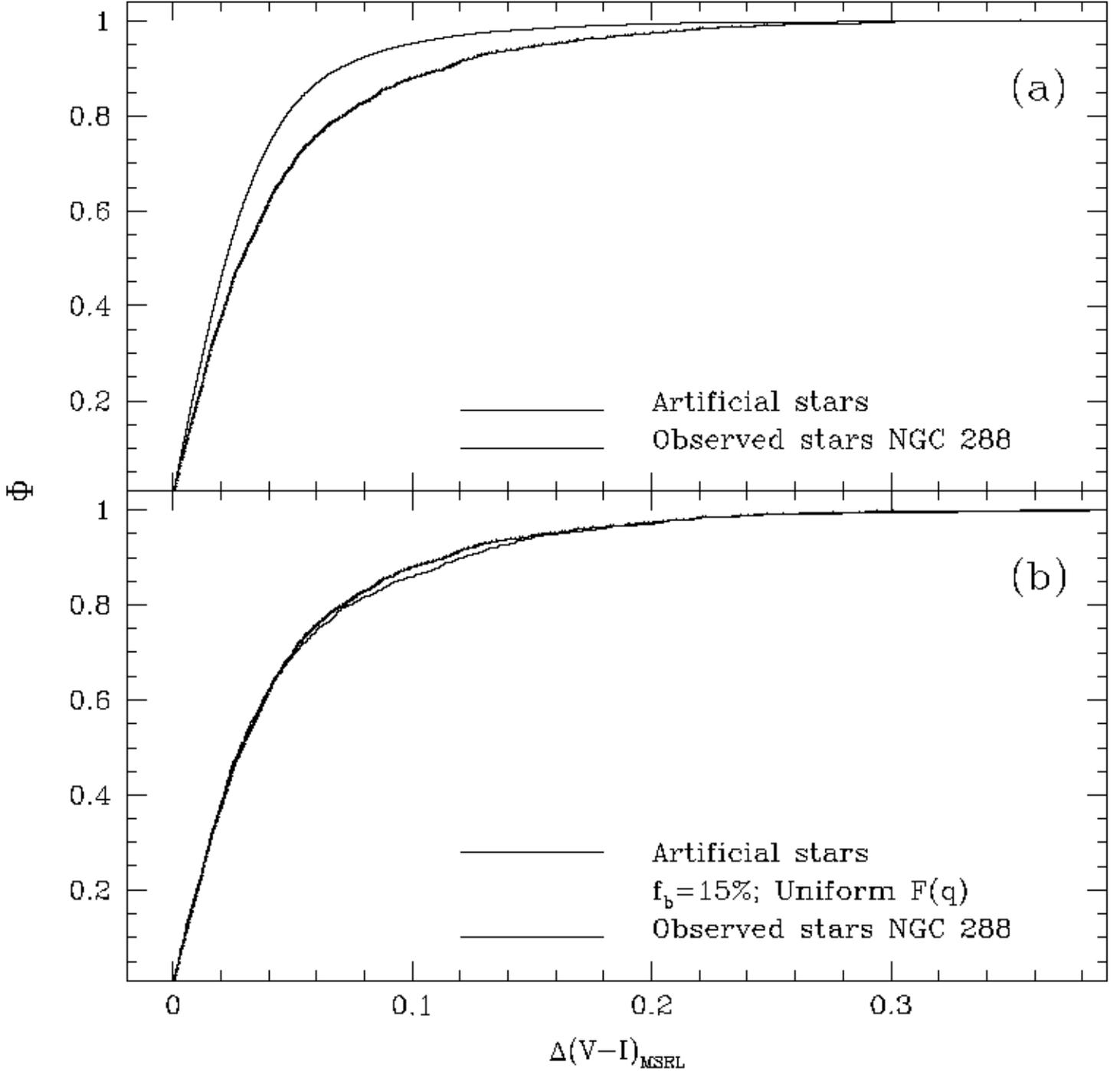}}
\caption{Panel ({\em a}): Comparison between the cumulative distributions of 
color deviations 
from the MSRL of the observed sample (thick line) and of a large subsample of
artificial stars (thin line). The color deviations are only for stars in the
range $20 \le V \le 23$ (see \S3.). Note that the observed distribution is
significantly skewed toward larger color deviations with respect to the
artificial star distribution that, by definition, contains no binaries.
The probability that the compared samples are drawn from the same parent
population is $P < 10^{-6} \%$.
Panel ({\em b}): a ``good fit'' case is also shown for comparison. The observed
sample is compared to a realization of a synthetic sample of the same dimension
with a binary fraction of $15 \%$ drawn from a uniform distribution of mass
ratios. The probability that the compared samples are drawn from the same parent
population is $P = 84 \%$.}
\end{figure*}

The results for all of the simulated samples are shown in
Fig.~12. Each of the small dots (in the vertical columns) shows the
probability that a given simulated sample is extracted from the same
parent population of the observed sample as a function of the $f_b$ of
the simulated sample. The different panels of Fig.~12 show the results
obtained assuming different $F(q)$ (from top to bottom: PLMR, PHMR,
Natural and Uniform). There are 100 of these for each $f_b$. The
open circles give the average probability.  The average points have
been connected by a continuous line as an aid for the eye and as a
first order interpolation between the obtained estimates.

\begin{figure*}
\figurenum{12}
\centerline{\psfig{figure=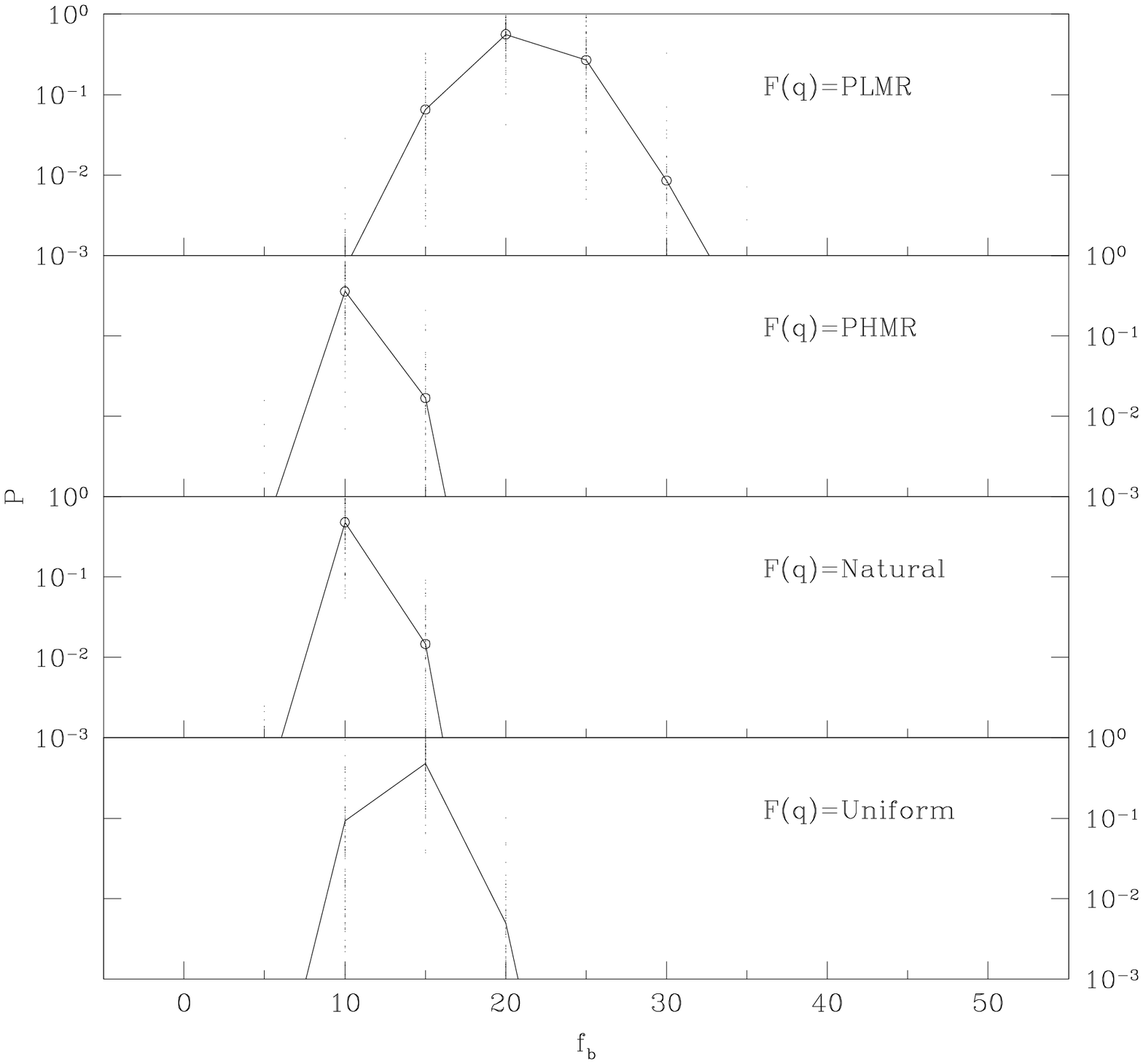}}
\caption{Estimates of the binary fraction in the whole observed sample
(Int+Ext), for the four $F(q)$ we considered. In each panel the
probability that the actual simulated sample is drawn from the same
parent population as the observed sample (according to a KS test) is
plotted versus the binary fraction introduced in the simulated
samples.  Each small dot represents one of the 100 random realizations
of the simulated sample at a given [$f_b,F(q)$] pair. The open
circles are the average probability of the 100 realizations of
simulated samples. A line connecting the open circles has been drawn
as an aid for the eye. The figure has been arranged in a way to be as
similar as possible to the corresponding plots by RB97.}
\end{figure*} 

There are many features worthy of comment in Fig.~12:

\begin{enumerate}

\item Independently of the assumed $F(q)$, the observed distribution of color
deviations is strongly incompatible with a binary fraction lower than $5$ \%.

\item Compatibility with observations (at least marginal) can be
obtained for 5 \% $\le f_b \le 35$ \%, depending of the adopted
$F(q)$.  However if we consider [$F(q),~f_b$] pairs which yield at
least one case in which $P > 1$ \%, the compatibility with $f_b=5$ \%
remains only for the PHMR case, and even then for only one realization
out of one hundred. On the other hand, the compatibility with $f_b=35$
\% is excluded in all cases and $f_b=30 $ \% is only allowed for the
PLMR. Hence, independently of the assumed $F(q)$, an appreciable
compatibility (i.e., $P>1$ \% in a non-negligible number of cases) is
reached only for $5$ \% $< f_b \le 30$ \%.

\item The {\em most probable} $f_b$ ranges from 10 to 20\% depending on the
assumed $F(q)$. 

\item As expected the PLMR case provides both the highest $f_b$
estimate and the widest range of compatibility. This is because in
this case most of the binaries are hidden near the MSRL and the
observed SMS would be a minor component of the whole population.

\item The estimates obtained assuming a PHMR, Natural or Uniform
distribution of mass ratios are rather similar. If these $F(q)$ are
considered, the $P> 1$ \% compatibility range is $5$ \% $< f_b \le 20$
\%.

\end{enumerate}   

The global field covers a significant fraction of the cluster within
2\,$r_h$. Thus, we consider the above estimates as fairly representative
of the whole population in NGC 288. Therefore, we conclude that {the
global binary fraction in NGC 288 is 5 \% $\le f_b \le 30$ \%},
independently of the assumed $F(q)$, while recalling that the range
giving the highest degree of compatibility with observations is $10 \%
\le f_b \le 20$ \%. This result is in good agreement with the lower
limit $f_b$ derived by \citet{bol92}.

\subsection{Radial Segregation}

A very basic prediction of the theory of dynamical evolution of
globular cluster is that such collisional systems tend to energy
equipartition which in turn generates mass segregation. The heavier
objects sink toward the central region of the system on timescales
comparable with the relaxation time \citep{mh97}.  The evolutionary
rate driven by two-body encounters is quite low in NGC 288 because of
its low stellar density. However, the cluster has an age comparable
with the Hubble time ($\ge 10$ Gyr) which is much longer than its
relaxation time (either central or median $t_{rc}\sim t_{rh} \sim 1$
Gyr; for either definitions and estimates see Djorgovski, 1993).  Thus,
the cluster should have dynamically relaxed long ago, and the
binary systems, that are heavier than single stars in average, are
expected to be significantly concentrated in the inner regions.

To verify this we considered separately the Int and Ext sample to
provide an estimate of $f_b$ in the two regions at different radial
distances from the cluster center. We take advantage of the lucky
circumstance that, with good approximation, the Int field samples the
region within $1 r_h$ from the center, while the Ext samples the
region between $1\,r_h$ and $2\,r_h$, so the comparison of the Int and
Ext fields has a well defined physical meaning.

In principle, one might desire to divide the sample in many radial
annuli, providing an estimate of $f_b$ in each annulus and obtaining a
radial profile of the binary fraction. In practice, this approach
turns out to be unviable because the method needs large samples to
provide significant estimates (see also RB97), so each subdivision of
the sample would reduce the constraining power of the test.

\begin{figure*}
\figurenum{13}
\centerline{\psfig{figure=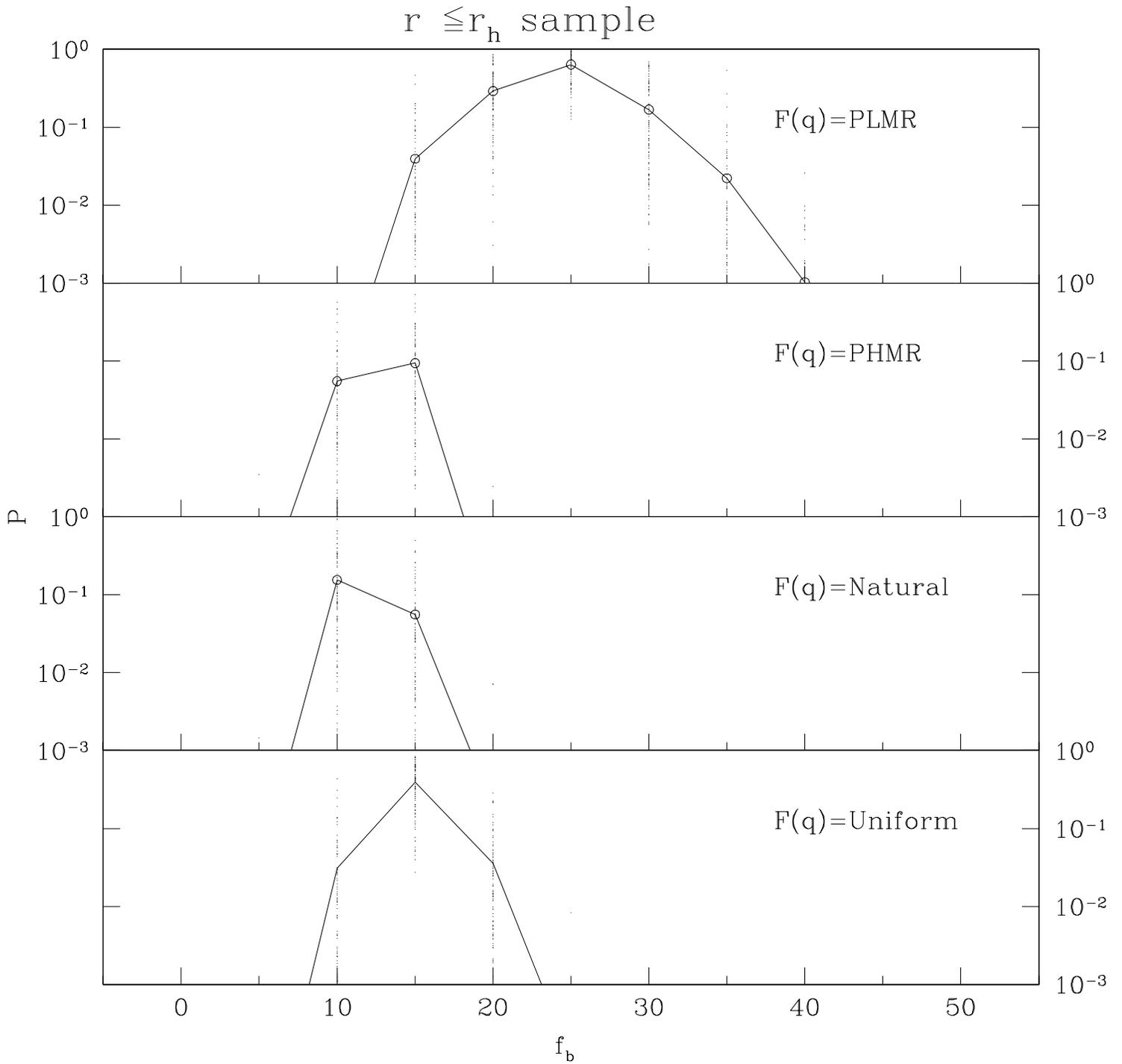}}
\caption{Estimates of the binary fraction in the Int sample, approximately
enclosed in the region within 1\,$r_h$ from the cluster center (see Fig.~1). 
The arrangement and symbols are the same as Fig.~12}
\end{figure*}

\begin{figure*}
\figurenum{14}
\centerline{\psfig{figure=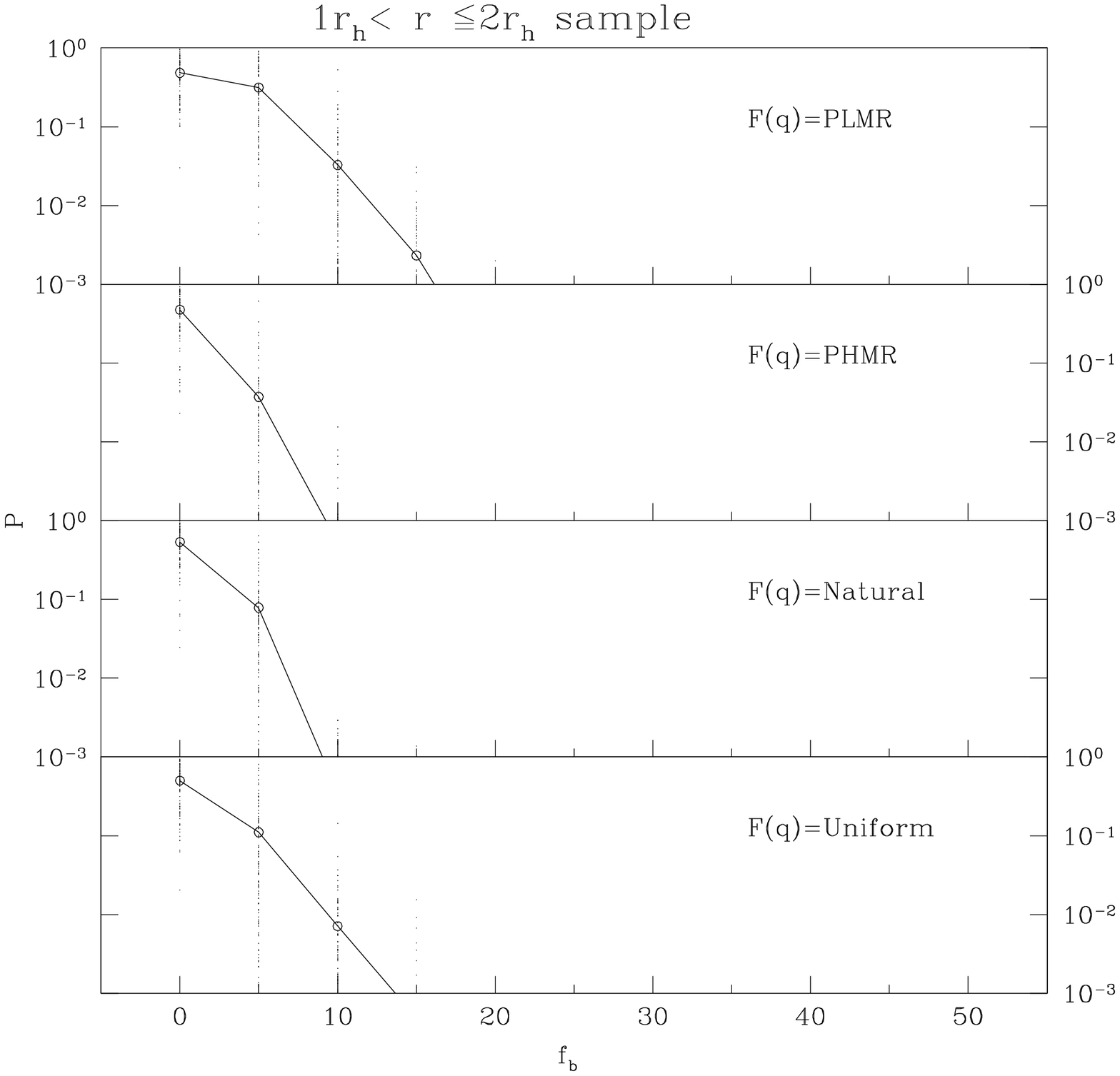}}
\caption{Estimates of the binary fraction in the Ext sample, approximately
enclosed in the region between 1 and 2\,$r_h$ from the cluster center 
(see Fig.~1). The arrangement and symbols are the same as Fig.~12 and 13.
Note that $f_b=0$ is by far the most likely binary fraction independently of the
assumed $F(q)$.}
\end{figure*} 

In Fig.~13 the $f_b$ estimates for the Int region are presented. A direct
comparison with Fig.~12 makes it immediately evident that the ranges of
compatibility have shifted toward higher $f_b$ values, independently of the
adopted $F(q)$. The results are no longer compatible with $f_b=5$ \%. The broad
range of compatibility is 8 \% $< f_b \le 38$ \%, while the most probable
$f_b$ values ranges from 10 \% $ to 25 $ \%, depending on the $F(q)$ considered.
These results suggest that {\em the binary fraction within $1\,r_h$ is
higher than in the global sample}.

Fig.~14 shows the results obtained in the Ext region. Independently of
the considered $F(q)$ {\em the most probable binary fraction is 0 and
in any case $f_b\le$ 10 \%} ($P> 1$ \% compatibility range).  Thus,
there is a strong indication of a significant difference in binary
fraction between the two considered radial zones. Furthermore, it can
be concluded that {\em most of the binary systems in NGC 288 reside
within $1\, r_h$ from the cluster center}, a clear effect of the mass
segregation.

It is worth emphasizing that the above results cannot be achieved by a
direct comparison of the $\Delta (V-I)_{MSRL}$ distributions of the
Int and Ext samples, since they suffer from different degrees of
crowding and, by consequence, have different photometric errors and
rate of blending. It is expected that the stars in the Int sample
have, on average, larger $\Delta (V-I)_{MSRL}$ with respect to the Ext
ones. It was essential to determine if the difference induced by
photometric errors and blendings was sufficient to explain the
observed difference in the distribution of color deviations and to
correctly quantify the effect of these factors. This was exactly the
aim of the adopted technique, without which neither absolute nor
differential $f_b$ estimates can be reliably obtained.

\subsection{The Nature and Evolution of Binaries in NGC 288}

Note that a binary fraction $f_b=10$ \%, a value near the lower limit
for the region within the half light radius, implies that 18 \% of
the cluster stars are bound in binary systems. Since the evolution of
such stars may be somehow influenced by the presence of a companion,
it is evident that a moderate binary fraction can have a
significant impact on the production of anomalous populations in the
cluster (BSS, sdB, etc).

With simple and basic theoretical arguments it is possible to derive
some useful (though statistical in nature) constraints on the origin of
the binary population of NGC 288. There are three possible origins for
a binary in a globular cluster: (a) the member stars were
gravitationally bound at their birth , i.e., the system is {\em
primordial}, (b) occasionally three single stars can have a close
encounter leaving two of them gravitationally bound, the third stars
having gained the excess energy, i.e., a {\em three body encounter},
or (c) during a very close encounter, ``two unbound stars can divert
enough orbital energy in the form of stellar oscillation that the pair
becomes bound'' \citep{hual}, i.e. a {\em tidal capture}.

According to \citet{bt94} the number of binaries formed by three body
encounters in a cluster having $N$ member stars is $\sim 0.1/(N \ln N)$
per relaxation time.  The predicted number of such systems formed over the
whole lifetime of NGC 288 is $\sim 10^{-7}$, under the very
conservative assumption that all members have mass $1\, M_{\odot}$ and
adopting the total mass estimate by \citet{pm93}, i.e.
$M=10^{4.9} M_{\odot}$.  
The rate of tidal
capture can be roughly estimated with the analytical formula provided
by \citet{lo86}. Under the most conservative assumptions it turns out
that the number of tidal captures that may have occurred in NGC 288 in
its whole lifetime is $< 10$. Thus, it can be concluded that {\em the
overwhelming majority of the binary systems in NGC 288 have a
primordial origin}.

Binary systems in a globular cluster may be broadly classified
according to the ratio between their intrinsic energy and the typical
kinetic energy of the cluster stars. Binaries are said to be {\em
hard} or {\em soft} if their energy is, respectively, larger or lower
than this value \citep{hual}.  As a general statistical rule,
encounters between a single star and a binary results in an hardening
for hard system (the unbound star is accelerated and the binary
becomes more tightly bound) while soft binaries are softened \cite[the
unbound star is decelerated and the binary becomes less tightly bound;
see][]{bt94,mh97}.  The average result of the evolution driven by
single star-binary encounters is that hard binaries increase their
energy reservoir and decrease their cross section for close
encounters, while soft binaries increase their cross section becoming
less and less bound.  Assuming an average mass of 0.2$\,M_{\odot}$ for
the members of NGC 288, according to \citet{krou}, it turns out that
{\em all} of the binary systems formed by two MS stars and with
semimajor axis larger than $\sim 5$ AU are
{\em soft}.  From
eq. 8 of \citet{hual} one finds that all of the primordial binaries
with semimajor axis larger than $\sim 7$ AU are likely to have had at
least one close encounter in their lifetime. Thus, it can be concluded
that the population of soft binaries in NGC 288 has significantly
evolved because of encounters with single stars, and probably most of
the original soft binaries have been destroyed. Finally, if one
considers only the population of hard binaries, neither
binary-single star nor binary-binary encounters are expected to
have had a large impact on the evolution of such 
systems\footnote{According to the rates of binary-binary encounters provided by
\citet{leo89} and assuming $f_b=0.2$, in line with our results, the fraction of
hard binary systems that are expected to have a close encounter during the
lifetime of the cluster is less than $10 \%$. While it seems unlikely that
binary-binary encounters do affect significantly the evolution of the binary
population of NGC~288 as a whole, the mechanism may be relevant as a channel 
for the production of blue stragglers (see \S5.0.1).}.

To summarize these points:

\begin{itemize}

\item the population of binaries in NGC 288 is largely dominated by {\em
primordial} systems;

\item the present day population is likely to be mainly composed of hard
binaries whose rate of evolution due to encounters is not great. 

\end{itemize}

It has to be noted that the above conclusions are based on the hypothesis 
that the stellar density in NGC~288 was not significantly higher in the past. 
This possibility is shortly discussed in \S 5.3.1.
    
\section{Blue Stragglers}

Blue Straggler Stars (BSS) are hydrogen burning stars that presently
have a mass higher than at their birth. This situation is thought to
arise either via the collision and merger of two single stars ({\em
ss} collision; a mechanism that may be efficient in the central
regions of the densest globulars), or via the mass transfer and/or
coalescence between the members of a binary system \cite[see][and
references therein]{stry,fp92,bai95,mat96a,ata97}. The coalescence
between the members of a binary system may be favored by
binary-binary encounters, an occurrence that greatly enhances the
probability of a physical collision between two of the member stars
\citep[{\em bb} collision, see][]{leo89}.  While there are only
indirect clues supporting the collisional scenario, the link between
(at least some) BSS and binary systems is firmly established
\citep{mat96b}. However, it seems probable that either single star and
binary mechanisms may be responsible of the formation of BSS in
different conditions, even in different regions within the same
cluster or {\em at different epochs} of the evolution of a cluster
\cite[see][]{fp92,fer93,fer95,fer97,fer99b,sig94,sills98}.

The presence of Blue Stragglers in NGC 288 was first noted by
\citet{alc80} and by \citet{buo84a,buo84b}.  \citet{bol92} found a
remarkable population of such stars and showed that BSS in NGC 288 are
more centrally concentrated than SGB and RGB stars of comparable
magnitude. This trend is observed in many other globulars and usually
interpreted as due to the settling of the more massive BSS to the
inner part of the clusters \cite[however the interpretation of the
radial distribution of BSS is not always so straighforward,
see][]{fer93,bai95}.

\subsubsection{BSS from binary-binary collisions in NGC~288}

Before proceeding in the analysis, we estimate the efficiency 
of {\em bb} encounters in NGC~288 to
check if this mechanism may be relevant for the production of BSS
stars in this cluster \citep{leo89,leof}.

According to Eq. 14 of \citet{leo89}, adopting the structural and kinematical
parameters from \citet[][$r_c$]{dj93} and \citet[][$M/M_{\odot},\sigma_0$]{pm93},
and assuming an average stellar mass of 0.2 $M_{\odot}$ as in \S4.2
\citep{krou}, the number of {\em bb} collisions\footnote{In the context of
binary-binary interactions {\em encounter} and {\em collision} are considered
synonyms, since we are dealing with encounters sufficiently close to
significantly alter the original status of the binaries involved. We are
referring to the actual collision between two stars as to ``physical collision''
or ``{\em ss} collision''.}
per gyr ($N_{bb}$ [Gyr$^{-1}$]) in NGC~288 is:

$$N_{bb} \simeq 387.6 a f_b^2$$

\noindent where $a$ is the semimajor axis of the considered binaries in AU. 
If only
hard binaries are taken into account ($a \le 5$ AU; see \S4.2) and a binary
fraction $f_b = 0.2$ is assumed (quite likely for the cluster core) $\sim 80$ 
{\em bb} collisions per gyr are expected. It is not clear how many of such 
encounters may ultimately lead to the physical collision between two
members of the involved systems. However it is clear that {\em bb} encounters
are a viable mechanism for the production of (at least) part of the BSS in
NGC~288, unless the fraction of {\em bb} encounters producing a physical
collision is significantly lower than ${1\over{100}}$.

\subsection{BSS in the UV: Specific Frequency}

A clear sequence of BSS is evident in the CMD of the Int sample in
Fig.~1, while the Ext sample hosts at most {\em one} clear BSS
candidate (see \S5.2).  We will discuss the radial distribution later
on, since it is safer to assess the selection criteria of candidate
BSS before. In Fig.~15 the BSS candidates are selected in the
($m_{F255W}$, $m_{F255W}-m_{F336W}$) CMD according to the criteria
introduced by \citet{fer97,fer99b}. In this plane the hottest sources
are especially obvious and the BSS sequence stands out as an almost
vertical plume at $16.50 \le m_{F255W} \le 18.65$ and $-1.0\le
m_{F255W}-m_{F336W}\le 0.4$.  Furthermore, adopting this
selection criterion allows a direct comparison with the clusters
studied by Ferraro and collaborators \cite[see][and references
therein]{fer00,bm00}. The F255W observations were available only for
the Int field, thus the CMD presented in Fig.~1 refers to the region
within $\sim 1\, r_h$ from the cluster center. The derived BSS specific
frequency $F_{BSS}$, defined as the ratio between the number of BSS
($N_{BSS}$) and the number of HB stars ($N_{HB}$) observed in the same
given field, is $F_{BSS}=1.35 \pm 0.49$, where the error is the
Poisson noise on the star counts. The specific frequency in the very
central region $r\le 1\, r_c$ is $F_{BSS}=1.45 \pm 0.59$, while in the
annulus $1 r_c< r\le 2 r_c$ it is $F_{BSS}=1.00 \pm 0.82$, hinting
that the BSS population is more centrally concentrated than HB stars.
It is very interesting to note that NGC 288 has a BSS frequency
significantly higher than all the other clusters for which $F_{BSS}$
has been measured with this technique, i.e. M13, M3, M92, M30 all
having $F_{BSS}\le 0.67$ \cite[see Table~1 in][and references
therein]{bm00}.  The only exception is the core-collapsing cluster
M80 in which \cite{fer99b} discovered an exceptionally abundant and
extremely concentrated population of BSS (305 stars). M80 has a
global frequency $F_{BSS}\simeq 1.0$, and $F_{BSS}\simeq 1.7$ in the
core, very similar to NGC~288. \cite{fer99b} argued that during the
phase of core collapse the extreme stellar density occurring in the
core of M80 boosted the star-to-star collision rate, thus providing
a very efficient mechanism for the production of collisional BSS.
Since, as said, {\em ss} collisions are quite rare in the loose NGC~288, the
observed {\em exceptionally high} specific frequency suggests that
{\em the formation of BSS via binary evolution} in low density
clusters can be as efficient as {\em the ss collisional mechanism} in
very dense ones. The possible consequences of this result will be
discussed in \S6 
\citep[see also][]
{rms,nh87,nc89,leo89,leof,stry}.  

Nearly at the center of Fig.~15, at $m_{F255W}\simeq 16.3$ and
$m_{F255W}-m_{F336W}\simeq 0.6$, one can see a HB star much cooler
than other stars in this phase. The same star stands out very clearly
in the $(V,~V-I)$ CMD, at $V\simeq 15$ and $V-I\simeq 0.8$, in the
panel ({\em a}) of Fig.~2. The position of this star in the CMD is
consistent with the hypothesis that it is an Evolved BSS \cite[E-BSS,
see][]{fer99b,bm00,fp92}, i.e. a BSS in its Helium burning phase.

\begin{figure*}
\figurenum{15}
\centerline{\psfig{figure=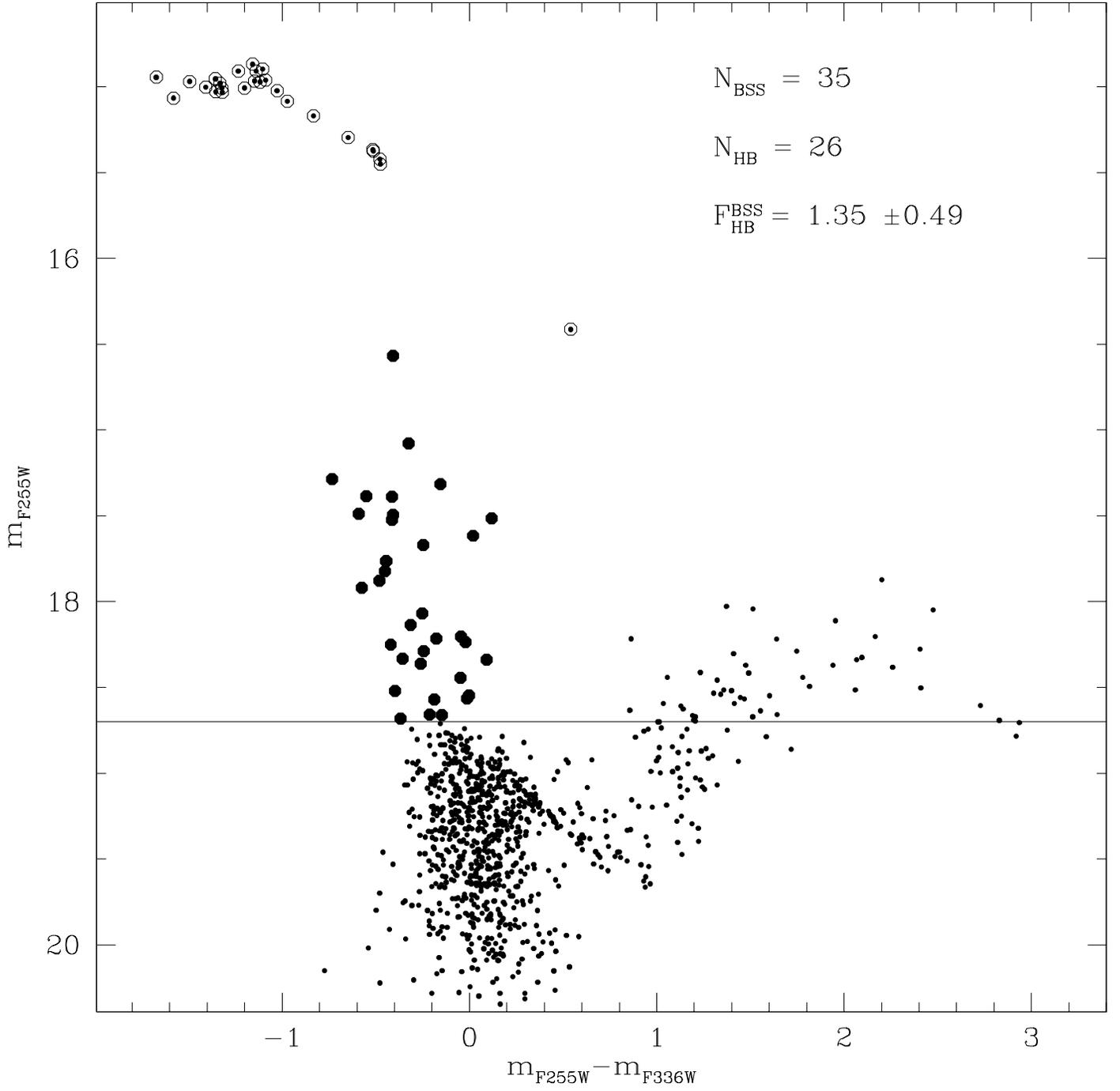}}
\caption{Blue Stragglers selected in the 
($m_{F255W}$, $m_{F255W}-m_{F336W}$) CMD. The F255W observations are available
only for the Int field.}
\end{figure*}

\subsection{BSS in the $V,~V-I$ Plane}

The selection criteria used above has been very useful in making
strictly homogeneous comparisons with the BSS population of other
clusters. However, to study the evolutionary status of our sample of
BSS we must adopt a selection criterion, based on the ($V,~V-I$)
CMD. Any such BSS sample may be somewhat contaminated by non-genuine
BSS. In the vicinity of the upper MS and SGB sequences, blended stars
and/or real detached binaries may be impossible to distinguish from
{\em bona fide} BSS. For NGC~288, we can take advantage of our
extensive set of {\em realistic} (with correct {\em colors})
artificial stars experiments to select a {\em pure} sample of BSS.  In
the panel ({\em a}) of Fig.~16 we show the CMD of a large ($\sim
200,000$ stars) randomly extracted subsample of artificial stars, in
the region where the observed candidate BSS lie. The large number of
artificial stars ensures that many blendings are included in the
sample. Indeed, a number of stars are found outside the narrow band
around the input ridge line. In particular, some blending between MS
and SGB stars populates a small region to the blue of the SGB. We
defined a selection box (the irregular polygon delimited by the thick
line) as a region of the $V$, $V-I$ plane that {\em is not
contaminated by blended sources}.  Note that this approach is
extremely conservative in avoiding spurious BSS. The number of stars
found between the red edge of the selection box and the single-star
SGB sequence is comparable in the artificial star CMD and in the
observed one (Fig.~16, panel ({\em b})), while the former sample is
$\sim 25$ times larger than the observed sample. This means that {\em
most} of the observed stars in that region of the CMD are {\em
genuine} BSS or detached binary systems, and only a small minority
must be attributed to blendings. (Note, for instance, the
concentration of observed stars to the blue of the MSTO, between
$V\simeq 18.8$ and $V\simeq 19.6$ and compare with the same region in
the artificial star CMD). Still, we chose not to include these stars
to ensure that {\em the contamination by blendings and/or detached
binaries is certainly null}, and that our sample includes only {\em
bona fide BSS}. This conservative approach causes us to lose roughly
20\% of the potential BSS sample based on the fact that 28 out of the
35 UV selected BSS candidates in Fig.~15 also fall in the $V,~V-I$
selection box adopted here.

\begin{figure*}
\figurenum{16}
\centerline{\psfig{figure=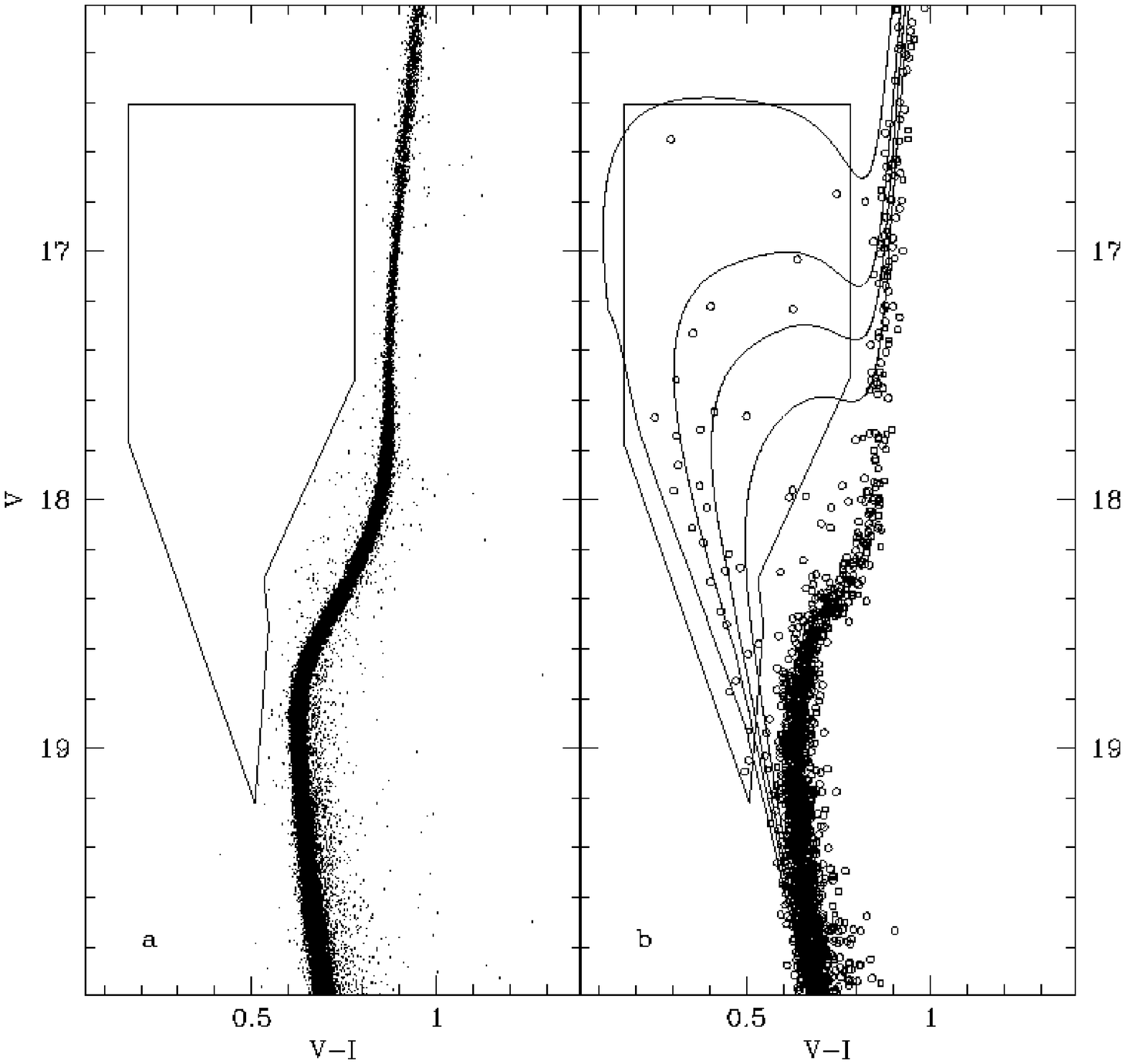}}
\caption{Blue Stragglers selected in the ($V$, $V-I$) CMD. Panel ({\em a}):
$\sim 200,000$ artificial stars and the {\em uncontaminated} selection box (see
text). Panel ({\em b}): the selection box is superimposed to the observed CMD
(circles: stars of the Int sample; squares: stars of the Ext sample). Four
isochrones ($Y=0.23$; $Z=10^{-3}$) from the set by \citet{cast}. From left to
right: Age = 2 Gyr, $M_{TO} = 1.33\, M_{\odot}$; Age = 3.5 Gyr, 
$M_{TO} = 1.13\, M_{\odot}$; Age = 4.5 Gyr, $M_{TO} =\, 1.05 M_{\odot}$; 
Age = 6 Gyr, $M_{TO} = 0.96\, M_{\odot}$.   }
\end{figure*}

In the panel ({\em a}) of Fig.~16, the selection box is superimposed to
the observed CMD (circles: Int sample; squares: Ext sample). The selected
BSS sample contains 33 stars, just one of them belonging to the Ext
sample\footnote{There are four BSS selected in the $(V,~V-I)$ plane that
are not included in the UV sample, because they are too faint or too cold
(or a combination of the two) to be isolated in the ($m_{F255W}$,
$m_{F255W}-m_{F336W}$) plane.  Two of them are the faintest sources in the
lower corner of the selecting box of Fig.~16b, at $V > 19$. The third is a
faint star near the red edge of the box at $V\simeq 18.6$ and $V-I \simeq
0.6$. The last is the bright yellow straggler at $V\simeq 16.8$ and $V-I
\simeq 0.7$, that is the coolest of the stars selected in Fig.~16b. This
star pass the UV selection criterion in magnitude, having $m_{F255W}
\simeq 18.2$, but not in color, with $m_{F255W}-m_{F336W}\simeq 0.9$.
However it stands clearly in the appropriate position of a cool yellow
straggler also in the UV CMD (see Fig.~15).}

Four isochrones of the appropriate chemical
composition ($Y=0.23$; $Z=10^{-3}$) from the set by \citet{cast} are
also shown in the plot, assuming the reddening and distance modulus
listed by \citet{fer99a}. The faintest and reddest isochrone has an
age of 6 Gyr, the mass of the stars at the TO is $M_{TO}=0.96\,
M_{\odot}$; then, going toward bluer color and younger ages, age = 4.5
Gyr, $M_{TO} =\, 1.05\, M_{\odot}$; age = 3.5 Gyr, $M_{TO} =\, 1.13
M_{\odot}$; and finally age = 2 Gyr, $M_{TO} = 1.33\, M_{\odot}$. The
reported isochrones encompass the whole distribution of BSS, thus
constraining the range of masses covered by the selected BSS to $0.9\,
M_{\odot} \le M_{BSS} \le 1.4\, M_{\odot}$.  Blue Stragglers with
$M_{BSS} \le 0.8\, M_{\odot}$, i.e. formed by the merging of low mass MS
stars, are still hidden in the single-star Main Sequence.

 There are some BSS that cannot lie on the MS of any reasonable
isochrone (at $V<17.4$ and $V-I>0.6$), but which can be well fitted by
the SGB sequences. Given our stringent selection criteria, there may
be very little doubt that these stars have {\em evolved} to the
thinning hydrogen burning shell phase and are rapidly moving to the
base of the RGB, i.e., they are {\em Yellow Stragglers}
\citep[YS;][]{hual,port97a,port97b}.

\subsection{Evolution of the BSS}

\subsubsection{The Conventional Merger Scenario and BSS Formation Rate}

Most modern BSS formation scenarios have the implicit hypothesis that
there is no preferred mass ratio for the stars that merge to form a
new BSS. For instance, in the simulations by \citet{sills99,sills00}
the merging stars are drawn at random from a properly choosen IMF of
single stars. Furthermore, Sills' models assume that {\em all} BSS are
the final result of a {\em merging} between two stars.  Given this,
simple considerations about the stellar lifetimes predict that if the
rate of formation of BSS were constant over the lifetime of a
globular, or at least from a given epoch in the past to the present
day, the number of observed BSS must increase toward the red-faint
part of the sequence, i.e., near the single-star MS and SGB. The
prediction is quantitatively confirmed by detailed models of BSS
populations \citep{sills98,sills99,sills00}.  This is certainly not
the case of NGC~288. Even in the region between the 2 Gyr and the 6
Gyr isochrones the number of BSS decreases toward the redder-fainter
direction.  There are 13 BSS between the 2 Gyr and the 3.5 Gyr
isochrones, 9 between the 3.5 Gyr and 4.5 Gyr isochrones and 5 between
the 4.5 Gyr and 6 Gyr isochrones. Taking the counts between 2 Gyr and
the 3.5 Gyr isochrones as normalization point and assuming uniform
rate of BSS production, crude estimates based on lifetimes would
predict $\sim 19$ BSS in the region between the 3.5 Gyr and 4.5 Gyr
isochrones and $\sim 24$ between the 4.5 Gyr and 6 Gyr isochrones. It
can be concluded that {\em the observed distribution of BSS in the CMD
is not consistent with a constant rate of production over a long time
} ($\sim 5$ Gyr or larger). 

The BSS in our sample lie in a remarkably narrow band,
significantly clustered around the 3.5 Gyr isochrone.  
Indeed, {\em the observed
distribution of BSS in NGC~288 resembles that predicted by
BSS evolutionary models with a short 1--2 Gyr burst of BSS
formation which occurred 1--4 Gyr ago,}
\cite[see  in particular Fig.~1 and Fig.~3 of ][]{sills00}.
       
A BSS formation rate with significant variations with time has been
recently invoked to explain the BSS distribution in M80 \citep{fer99b}
and in 47~Tuc \citep{sills00}. A possible explanation for the 
enhancement of the BSS production rate that appear to have occured in NGC~288
may be related to the process of mass segregation that, during the lifetime of
the cluster, has collected more and more binaries in the cluster core, thus
enhancing the rate of {\em bb} collisions. However, according to the present day
relaxation time, the segregation of binaries into the cluster core should have
occurred long time ago, at most a few $t_{rh} \sim 1$ gyr after the birth of the 
cluster. 

The conclusion that phenomena driven by collisions are highly inefficient in
NGC~288 (\S 4.2.) has been drawn on the basis of the {\em presently} observed
structural and dynamical conditions of this cluster. However such conditions may
have been different in the past.
It may be conceived that NGC~288 was significantly denser in the past and has 
been brought to the present status by the disruptive effects of disc and bulge 
shocks. If this were the case, it can be imagined that the peak of BSS
production rate coincided with the epoch of maximum contraction of the cluster,
during which BSS may have been efficiently produced by stellar collisions. The
subsequent expansion decreased the collision rate to the present level,
virtually stopping the production of collisional BSS.

While this framework provide an explanation for the deduced variation of
the BSS formation rate with time, there are reasons to regard it as unlikely.
In fact, to raise the collision rate at the required level it would have been
necessary for NGC~288 to reach central densities some 100 - 1000 times larger
than the present one. Since shocks fasten the internal evolution, the cluster
would probably reach the core collapse phase \citep{mh97,oleg}. Even if binaries
stopped the collapse, it would be very difficult to re-expand the cluster {\em
core} to the present status. While most of the cluster halo may have been torn
apart in the subsequent perigalactic passages, the dense core would have become
very resistent to shocks. Thus the cluster present-day density would have been
sinificantly higher than observed. On the other hand, if the cluster experienced
a shock so strong to alter its {\em whole} structure, this would finally lead to
its complete disruption. Since from its birth NGC~288 had $\sim 50$ perigalactic
passages, probably it would have not survived to the present day.

\subsubsection{An Alternative View: Evolutionary Mass Transfer}

Let's consider now the alternative hypothesis that the BSS population in 
NGC~288 is dominated by stars that increased their original mass via mass 
transfer from their primary \citep[][hereafter CJ84]{mccr64,cj84}.

In this case the rate of formation of BSS will be driven by the rate that
stars leave the main sequence.  At the age of NGC~288 0.8\msun\ stars
leave the main sequence and might begin to transfer mass on the lower RGB.
They might dump most of their mass down to some helium core, say 0.2\msun,
onto the secondary. If $F(q)$ is peaked for large $q$, say 0.6--1.0, then
the resulting star will have a mass 1.08--1.40\msun. Even for a natural
$F(q)$ only 18\% of binaries have $q< 0.4$ so 82\% of BSS would have
$M>0.92$\msun. As the mass of the secondary increases the Roche lobe moves
toward the primary. Mass transfer will take place more easily. Indeed
stable mass transfer can occur only in binary systems with mass ratio not
too far from unity.  For example CJ84 assume a critical mass ratio
$q_{\rm crit}\ge 0.4$ as the condition for stable mass transfer. With this
condition, the lowest possible mass for a present day forming BS in
NGC~288 would be 0.92\msun.  

In the scenario described above there is a strict lower limit for the
mass of a BSS and the formation of more massive BSS is clearly
promoted. In this case the observed BSS distribution may be
accomodated with a rate of production running at a pace set by the
evolutionary rate of cluster stars (e.g., see the predicted BSS
distribution in Fig.~2g by CJ84).  It is interesting to note that in
the CJ84 model (\# 12) that is most like a typical globular cluster, a
high specific frequency of BSS is (crudely) predicted, $F_{BSS} \sim
2.5$ within a factor 2 of what observed in NGC~288\footnote{At the
time of the CJ84 analysis the only globular in which BSS had been
discovered was M3. CJ84 argued that the presumed lack of BSS in other
globular clusters was due to a lower binary fraction in halo
population and to a high rate of binary destruction occurring in these
dense environments. Now we know that NGC~288 is not dense enough to
have its whole binary population destroyed, and in fact many binaries
and a high specific frequency of BSS are observed.}. Note that CJ84
assumed empirically-determined distributions of binary orbital periods
and mass ratios, thus the fraction of mass-transfer systems producing
BSS was not artificially boosted.

 The framework depicted above seems to provide a natural explanation
for the observed properties of the BSS population in NGC~288 without
invoking a burst of BSS formation. Still it is clear that
understanding the NGC~288 BSS population requires more work.
Spectroscopic follow-up of individual BSS might help to distinguish
between different scenarios, though the spectral signatures of the
various possible origins are also not well defined
\citep{stry,bai95,sills98}.

\section{Summary and Discussion}

We have measured the fraction of binary system ($f_b$) in the loose
globular cluster NGC~288 using the SMS method to identify binaries in
the $V,~V-I$ CMD. We employed a technique that accurately 
accounts for {\em all} the observational effects (observational scatter,
blending, etc., see RB97) that may affect the estimate of $f_b$.  This
is only the second measurement of this kind ever obtained for a
globular cluster, the first having been made by RB97 for NGC~6752.

We find that the observations are strongly incompatible with the
hypothesis $f_b = 0$ \%, independently of any assumption concerning
the distribution of mass ratios, $F(q)$, thus {\em the presence of
binary systems in NGC 288 is confirmed beyond any doubt} \cite[see
also][]{bol92}. We have estimated $f_b$ for various assumed $F(q)$,
and have found that, independently of the adopted $F(q)$, {\em the
observations are strongly incompatible with global binary fractions
lower than 5--7\% or higher than $\sim 30$ \%}.

The binary fraction in the region within the half-light radius ($r_h$)
is significantly larger than in the region outside this limit. For $r
< 1\, r_h$, 8 \% $< f_b \le 38$ \% and the most probable binary
fractions ranges from 10 \% to 25 \%, depending somewhat on the
assumed $F(q)$. On the other hand, for $r > 1\, r_h$, $f_b \le 10$\%
and the most probable binary fraction is $f_b = 0$ \%, independently
of the assumed $F(q)$.  Hence, {\em binary systems are much more
abundant in the inner regions of the cluster, a clear sign of the
occurrence of dynamical mass segregation}.

Simple dynamical arguments strongly suggest that {\em the large
majority of binary systems present in NGC~288 are of primordial
origin}, and that single star - single star collision processes are
highly inefficient in this cluster. Thus the large majority of the BSS
in NGC~288 must have a binary origin.  Despite that, {NGC~288 has a
very high Specific Frequency of BSS, comparable to or exceeding that
of much more dense clusters}.  Like the binaries, the BSS population
is centrally concentrated with virtually all the identified BSS lie
within $1\,r_h$ from the cluster center.

The selected BSS candidates appear to have masses between $\sim 0.9
M_{\odot}$ and $\sim 1.4 M_{\odot}$, and form a remarkably narrow and
well defined sequence in the $(V,~V-I)$ CMD. If the majority of BSS
have been produced by the merging of two binary member stars, then the
{\em BSS distribution is  not compatible with a BSS formation rate which
has been constant in time}. On the contrary, the existence of a
significant peak in the BSS formation rate in the recent past ($\sim
1$--4 Gyr ago and lasting $\sim 1$--2 Gyr) is suggested by the
comparison with theoretical models \citep{sills00}.  The
observations also seem compatible with a scenario in which most of
the BSS population of NGC~288 has been producted via the mass transfer
occurring in close binary systems.

A few {\em yellow stragglers} and one candidate E-BSS (in the Helium burning
phase) have also been identified.

\subsection{Binary Fraction: Comparisons with Other Systems}

The two globular clusters for which a robust estimates of the {\em
global} binary fraction have been obtained using the SMS technique,
i.e., NGC~6752 (RB97) and NGC~288 (this work) are very different in
terms of stellar density. The central density in NGC~6752 is $\sim
1000$ times that of NGC~288.  It is very likely that the past
dynamical evolution of the two clusters has been very
different. Despite of that, the present day observed binary fraction
(and binary distribution) is remarkably similar, at least at the
present level of accuracy. However, while many details of the radial
distribution of the binary populations are beyond the reach of the
present day techniques, a first comparison of their broad properties
is now possible.  For NGC~6752 $15 \% \le f_b \le 38 \%$ within $1\,
r_c$ from the cluster center and $f_b \le 16 \%$ in the outer
region. For NGC~288 8 \% $\le f_b \le 38$ \% within $1\, r_h \simeq 1.6\,
r_c $ and $f_b \le 10$ \% in the outer region. It may be presumed that
a broad upper limit is set by primordial conditions and by the
fact that in any globular cluster a significant number of binaries are
{\em soft} (and thus are rapidly destroyed).  Lower limits may be set
by the equilibrium between destruction and formation of binary
systems, somehow regulated by the density of the environment
\cite[since the efficiency of both mechanisms increase with stellar
density,] []{hual}.

It is also interesting to note that the global $f_b$ extrapolated from
spectroscopic binaries and/or eclipsing variable searches in globulars
is broadly constrained to be in the range 10 \% $ \le f_b \le 40 $ \%
\cite[see][and references therein]{mat96a}.  Hence, two decades after
the pioneering (and negative) results by \citet{gg}, a very different
picture of binaries in globular clusters seems firmly established:
{\em the binary fraction in globulars is not null (and likely larger
than $\sim 5-10$ \%), but is still significantly lower than in most of other
environments}, i.e., the local field, open clusters and star forming
regions, for which $f_b\ge50$ \% \citep{mat96a,math,duqma}.

\subsection{Blue Stragglers in Low Density Systems} 

One of the most interesting results of the present analysis is the demonstration
that {\em the formation of BSS via mass transfer/coalescence of primordial
binary systems may be as efficient as collisional mechanisms, occurring in the
most favorable environments} (see \S5). This is a further (and quantitative)
piece of evidence indicating that {\em
large populations of BSS may be produced in environments with remarkably 
low stellar density, if a sufficient reservoir of primordial binaries is
available} 
\citep[see also][and references therein]{nh87,nc89,leolin,leo93,mat96a,mat96b}. 

This statement may have important consequences in the interpretation
of the ubiquitous {``blue plumes''} observed above the main MSTO in
the CMD of many dwarf Spheroidal galaxies (dSphs) dominated by old
``globular-cluster-like'' populations \cite[as for instance Ursa
Minor, Draco, Sextans and Sagittarius, see][respectively, and
references therein]{mart01,cs86,mat91,sgr2}.  These sparsely populated
sequences are usually associated with recent, very small episodes of
star formation. This hypothesis is also supported by the presence of
stellar species thought to result from the evolution of stars with initial mass
larger than the typical TO mass of old population (1--$2\,M_{\odot}$
vs. $0.8\, M_{\odot}$), such as Anomalous Cepheids (AC) and/or bright
AGB stars \cite[see][and references therein]{mat98,dacosta}.  Even
though the blue plumes lie in the region of the CMD populated by BSS,
BSS are sometimes considered an unlikely explanation because of the very low
density environments.

The results presented here confirms that a dense environment is not 
a {\em conditio sine qua non} for the efficient production of
BSS. Further, there is no reason to believe that primordial binaries are
under-abundant in dSphs \citep[see also][and references therein]{leof,leo93}. 
Indeed, the few available estimates suggest
that (at least in Ursa Minor and Draco) the binary fraction may be
even larger than in the local field \citep{opa}. It is also important
to recall that once a star more massive than the typical old TO stars
is formed from a binary system (via mass transfer or coalescence), it
will follow the evolutionary path typical of its {\em new} mass, thus
possibly passing through the AC or bright-AGB phases \cite[as
succesfully demonstrated by][more than two decades ago]{rms}. There is
every reason to suspect a significant BSS population in dSphs, and
they remain a viable explanation for the blue plumes \citep[see, e.g.][]{grill}.

\acknowledgments 

The finacial support of the {\em Agenzia Spaziale Italiana} (ASI) is kindly 
acknowledged. RTR is
partially supported by NASA LTSA grant NAG 5-6403 and STScI grant
GO-8709.  Part of this work has been the subject of the Thesis of
Degree of L. Monaco (Dept. of Astronomy, Bologna University).  Part of
the data analysis has been performed using software developed by
P. Montegriffo at the Osservatorio Astronomico di Bologna.  This
research has made use of NASA's Astrophysics Data System Abstract
Service.


\clearpage

\begin{deluxetable}{crrrrrrrrrrr}
\tabletypesize{\scriptsize}
\tablecaption{Observational material. \label{tbl-1}}
\tablewidth{0pt}
\tablehead{
\colhead{Field} & \colhead{Filter}   & \colhead{$t_{exp}$ [s]}   &
\colhead{N}  
}
\startdata
Int  & F814W & 140 & 4    \\
Int  & F814W &  12 & 2    \\
Int  & F814W & 1.2 & 1    \\
Int  & F555W & 100 & 4    \\
Int  & F555W &  14 & 2    \\
Int  & F555W & 2.3 & 1    \\
Int  & F336W & 1700 & 1   \\
Int  & F336W & 1000 &  2  \\
Int  & F336W &   60 &  1  \\
Int  & F255W  &  350 & 2 \\
Ext  & F814W & 160   & 3  \\
Ext  & F814W & 40   & 1   \\
Ext  & F555W & 230  & 2   \\
Ext  & F555W & 200  & 1   \\
Ext  & F555W & 40   & 1   \\
\enddata

\tablecomments{The fourth column (N) reports the number of repeated exposures
taken in a given passband.}

\end{deluxetable}

\clearpage

\end{document}